# Enhancing Trustworthiness in ML-Based Network Intrusion Detection with Uncertainty Quantification

Jacopo Talpini,[1] Fabio Sartori,[1] and Marco Savi[1]

[1] *Department of Informatics, Systems and Communication, University of Milano-Bicocca, Milano, Italy*[*]

The evolution of Internet and its related communication technologies have consistently increased the risk of cyber-attacks. In this context, a crucial role is played by Intrusion Detection Systems (IDSs), which are security devices designed to identify and mitigate attacks to modern networks. Data-driven approaches based on Machine Learning (ML) have gained more and more popularity for executing the classification tasks required by signature-based IDSs. However, typical ML models adopted for this purpose do not properly take into account the uncertainty associated with their prediction. This poses significant challenges, as they tend to produce misleadingly high classification scores for both misclassified inputs and inputs belonging to unknown classes (e.g. novel attacks), limiting the trustworthiness of existing ML-based solutions. In this paper, we argue that ML-based IDSs should always provide accurate uncertainty quantification to avoid overconfident predictions. In fact, an uncertainty-aware classification would be beneficial to enhance closed-set classification performance, would make it possible to carry out Active Learning, and would help recognize inputs of unknown classes as truly unknowns, unlocking open-set classification capabilities and Out-of-Distribution (OoD) detection. To verify it, we compare various ML-based methods for uncertainty quantification and for open-set classification, either specifically designed for or tailored to the domain of network intrusion detection. Moreover, we develop a custom model based on Bayesian Neural Networks to ensure reliable uncertainty estimates and improve the OoD detection capabilities, thus showing how proper uncertainty quantification can be exploited to significantly enhance the trustworthiness of ML-based IDSs.

## I. INTRODUCTION

Network intrusions stand as a major scourge within modern communication networks. With the constant increase in the number and complexity of occurring incidents [19], it is fundamental to design and implement scrupulous detection strategies and robust counteraction measures to effectively identify and mitigate these threats. *Intrusion Detection Systems* (IDSs) are among the primary security measures in communication networks, with the aim of identifying attacks, unauthorized intrusions, as well as malicious activities [59]. The traditional approach for detecting intrusions relies on *knowledge-based* systems [26] but as long as networks rise in complexity they become more prone to errors [52][58]. As a consequence, data-driven approaches based on *Machine Learning* (ML) have been widely considered in recent years for the detection of attacks.

IDSs are generally divided into *signature-based* or *anomaly-based* [58]. The first category, also known as *misuse-based IDSs*, is based on pattern recognition, with the goal of comparing signatures of well-known attacks and benign traffic to the current network traffic patterns. In this context, supervised ML methods are promising tools to analyze network traffic and classify it as benign or as a particular intrusion [56]. On the other hand, anomaly-based methods rely on a model for the normal (i.e., benign) network traffic so that any pattern that deviates from the usual one is considered an intrusion. In contrast to the misuse-based IDSs, anomaly-based IDSs are able to detect also new types of attacks, but they typically suffer from a high rate of false positives [35].

ML-based IDSs have been extensively investigated in the literature particularly, as said before, in the context of signature-based intrusion detection. These systems have demonstrated great performance in terms of classification scores [35]. However, the vast majority of the proposed methods in literature for signature-based intrusion detection rely on the implicit assumption that all class labels are a-priori known. This means that they are designated to perform the classification in a *closed-set* setting where each input is assigned to a category belonging to the set of labels provided during the training phase. However, in practice, a ML model might be presented with input pertaining to classes never seen at training time [5]. This scenario is particularly compelling in the context of intrusion detection [56], since networks are subject to new attacks and the likelihood of being targeted by zero-day attacks is getting higher and higher, especially in modern network infrastructures [70].

Moreover, several studies (e.g. [45][6][23]) have shown that ML models, including Deep Learning ones, tend to produce, overconfident, arbitrarily high per-class classifi-cation scores not only to misclassified samples from known classes but also to *Out-of-Distribution* (OoD) instances, which may be related to unknown classes. This behaviour represents a severe issue for the deployment of a *trustworthy* ML-based IDS since we might expect that an IDS will have to face new kinds of attacks or variations of known attacks [56]. In fact, it would be desirable to rely on a model that allows network operators or companies offering an IDS service to assess more accurately the *uncertainty* associated with the chosen model's predictions, thus raising their *awareness* and allowing them to perform more informed risk evaluations, while taking the corresponding most appropriate countermeasures accordingly.

* {name.surname}@unimib.it

We thus argue that for safety-critical applications, such as intrusion detection, the adopted ML model should be characterized not only through the lens of classification performance (accuracy, precision, recall, etc.), but it should also:

1. Provide *truthful uncertainty quantification* on the predictions for *closed-set classification*, a crucial property to avoid making wrong and overconfident decisions when the outcome is too uncertain, to help in the context of risk decision making. Having reliable uncertainty estimates is valuable also for performing *Active Learning* [15], i.e., a process where a ML-based system could learn from small amounts of data, and choose by itself what data should be labelled by a domain expert. This is a crucial aspect for the training and deployment of ML models in storage and memory-constrained scenarios (e.g. Edge Computing [62]) or when the labelling process of the data is demanding as in network intrusion detection [5], where huge amounts of data can be extracted in real time by exchanged traffic flows.

2. Be able to *recognize as "truly unknowns" inputs belonging to unknown categories*. This can be done by adopting uncertainty-aware classifiers, which can perform at the same time the usual closed-set classification and also be able to detect OoD samples in an *open-set classification* setting (i.e., with also unknown classes at test time)[1].

The goal of this paper is to perform a critical comparison of *uncertainty-aware ML models* and *open-set classifiers* that could be used in the context of network intrusion detection, considering an uncertainty-unaware ML-based IDS as a baseline. Moreover, we propose a custom model, based on Bayesian Neural Networks (BNNs), which is designed to offer well-founded uncertainty estimates for the classification of known network intrusions while enhancing the detection of unknown traffic patterns, by reducing the number of false alarms. To this end, we designed a method to recalibrate the uncertainty predicted by a given trained BNN to enforce high uncertainty for inputs far away from the training data, without adding substantial computational overhead. Our illustrative experimental results are obtained on an open-source dataset [4], and show that the adoption of uncertainty-aware models based on Neural Networks (NNs), Bayesian Neural Networks and Random Forests (RF) is very beneficial in the considered context, as they are able *(i)* to perform truthful uncertainty classification in a closed-set scenario, *(ii)* while also supporting Active Learning for efficient data labelling, and *(iii)* to enhance OoD Detection with respect to existing ad-hoc open-set classifiers. Moreover, we show how the proposed model stands out for its ability in detecting OoD samples, by reducing the false positives, and by showing a higher robustness across different OoD experiments, in comparison to other state-of-the-art methods. With our work, we therefore pave the way towards the adoption of uncertainty-aware ML models in risk-sensitive applications, such as intrusion detection, where the problem of uncertainty quantification and OoD Detection is crucial and still in its infancy.

The remainder of the paper is organized as follows. Section II and Section III are devoted to introduce related works and methods. In particular, in Section II we discuss some relevant related works on open-set classification in the context of intrusion detection. In Section III, we revise the general approach for uncertainty quantification with a focus on Bayesian Neural Networks, given their principled uncertainty-aware nature, and OoD Detection. Section IV is dedicated to formulating the problem statement, presenting the considered state-of-the-art models, and introducing our proposed approach. In Section V, we describe the utilized dataset and how we carried out data preprocessing. In Section VI we provide and discuss the numerical results, and conclude the paper by highlighting the main takeaways and lessons learned in Section VII.

## II. RELATED WORKS

In recent years, data-driven approaches for developing signature-based IDSs have been extensively explored (e.g. [52][58][26]) considering different methods such as Random Forests, Support Vector Machines, Neural Networks or Clustering techniques. Various Machine Learning and especially Deep Learning models have emerged as promising data-driven methods with the capability to learn and extract meaningful patterns from network traffic, which can be beneficial for detecting security threats occurring in networked systems [4]. However, it is important to stress that the vast majority of these ML-based IDSs are tested in a *closed-set* scenario. For instance, [63][64] compare different classification algorithms for developing an IDS and, in general, the best performance is achieved by tree-based classifiers, like Random Forests and Multi-Layer

[1] In this paper, when referring to *OoD Detection*, we will always implicitly refer to the context of *open-set classification*. This means that the two terms can be used interchangeably, with the assumption that OoD samples belong to unknown classes.

Perceptrons (MLPs).

To the best of our knowledge, in the field of ML-based intrusion detection only few works have addressed the specific problem of open-set classification to enhance signature-based approaches, while the more general problem of *uncertainty quantification* (also beneficial for enhanced closed-set classification and for Active Learning) is still unexplored. In the following, we thus discuss relevant related works *(i)* on Active Learning based on uncertainty quantification and *(ii)* on open-set classification in the considered domain, while an exhaustive review of traditional ML models adopted by IDSs is beyond the scope of this paper.

In the realm of uncertainty quantification in support to *Active Learning* for intrusion detection only a few works can be found in literature: [12][25][7] exploit the total uncertainty (more details in Section III) of ML models, mainly Neural Networks, to acquire samples to label. We will show why this approach is sub-optimal and how an IDS may take advantage of a more appropriate uncertainty quantification, enhancing the trustworthiness and efficiency of such a process in a closed-set classification scenario.

On the other hand, the literature on the *open-set classification* problem for network intrusion detection is richer. As an early contribution, [33] proposed a hybrid IDS, which combined an anomaly detection module based on Spark ML and a signature-based detection module based on a Convolutional-LSTM network classifier. In this way, it is possible to improve the scalability of intrusion detection by combining an anomaly detection method with a closed-set classifier.

More recent and competitive works based on a single model rather than a hybrid system for tackling the open-set classification problem are [69] and [57]: this is the approach investigated in this paper. In [69] the authors propose the “Open Set Classification Network" (OCD), a Convolutional Neural Network trained using both fisher loss [68] and MMD loss [22]. The rationale is trying to learn an optimal feature representation in the hidden layers of the network so that feature representations within the same known class are close together, while the feature representations of the unknown class and the known class are as far apart as possible. For that purpose, the authors propose to synthesize samples of possible unknowns to ensure the second phenomenon during the training phase. While this approach may be intriguing, its drawback lies in the necessity for ad-hoc synthetic training data to simulate unknowns. In contrast, in this paper we focus on methods trained only on known kinds of attacks, as a common supervised classification task, without making any explicit assumption on the possible OoD inputs. We argue that, in general, it will not be possible or practical to make a model aware of all of the possible unknowns. As an alternative, it may be sufficient for the model to detect that an input is ambiguous or novel, and then to react in an appropriate way, or require the intervention of a human expert for taking a decision.

More recently, [57] proposed EFC, an Energy-based Flow Classifier able to tackle the open-set classification problem. It is a statistical model for finding a probability distribution characterizing the per-class flows, and then it calculates the flow energy to quantify how likely a flow belongs to a given probability distribution. So, if the energy is low for a given flow, it is more likely that it belongs to the set of flows that generated the posterior distribution (see Section III) for that class, while if the energy is above a certain threshold, it can be considered as an unknown. We decided to incorporate this method as a reference for our study due to its promising performance in terms of OoD detection with respect to previous models. Moreover, this approach stands out because it can achieve reliable results by utilizing only known classes as training data, eliminating the need of synthesizing unknown samples.

Last, it is possible to find a vast literature regarding the problem of *zero-day attack detection* (e.g., [24][3][29]). However it should be noted that most of these works tackle the problem of detecting unknown attacks as a pure anomaly-detection problem or as a binary classification problem (benign traffic vs. attack). On the other hand, here, we propose the adoption of an end-to-end approach, with a *single model* that can retain the classification performance (in terms of high detection accuracy and recall) of a pure signature-based IDS on known kinds of attacks, while adding the capability of detecting unknowns, an aspect typical of pure anomaly-based solutions. In essence, our analysis differs as we are addressing an open-set classification task: we argue that the development of an uncertainty-aware IDS is a mean for reaching this goal, in addition to the previously described advantages compared to usual classification methods.

## III. UNCERTAINTY QUANTIFICATION AND OOD DETECTION

In this Section we begin by presenting the concept of uncertainty quantification and Bayesian Neural Networks. Subsequently, we focus on the problem of Out-of-Distribution detection, discussing specialized methods tailored to address this specific problem and how uncertainty-aware models can effectively address this challenge.

### A. Uncertainty Quantification and Bayesian Neural Networks

A crucial aspect of a model trustworthiness is the quantitative assessment of the model's uncertainty about its own predictions. In general, the uncertainty in model predictions can be decomposed in *aleatoric uncertainty* and *model* (or *epistemic*) *uncertainty* [20]. Aleatoric uncertainty arises from the inherent randomness in the input data, whereas epistemic uncertainty from the lack of the model's knowledge. The latter may be due to samples either out of distribution or sparsely covered by the train-ing set. The *Bayesian Inference* theory [66] provides a framework for quantifying and decoupling aleatoric and epistemic uncertainty in a principled way: in the following, we will review some basic results related to this approach.

More specifically, we will refer to a *supervised classification* problem, in which a set $\mathcal{D}$ of $N$ input-output pairs $\mathcal{D} = \{(\boldsymbol{x}_i, \boldsymbol{y}_i)\}_{i=1}^N$ is given, and the aim is to define a parametric function (e.g. a Neural Network) that provides the conditional probability distribution $p(\boldsymbol{y}|\boldsymbol{x}, \boldsymbol{w})$ over $K$ classes, for a given input $\boldsymbol{x}$ and model parameters $\boldsymbol{w}$ [44]. A particular set of parameters $\hat{\boldsymbol{w}}$ is typically chosen during the training phase by minimizing a loss function (e.g. the negative log-likelihood) exploiting a given dataset $\mathcal{D}$, and used to make predictions. In order to provide a proper probability distribution over $K$ classes, traditional NNs employ the softmax activation function in the output layer [44], which is often erroneously interpreted as model confidence on the classification predictions [20][44]. This pitfall is essentially due to the fact that this approach can not capture properly the epistemic uncertainty of the model [43], which is crucial especially for safety-critical applications.

However, it is possible to tackle this issue in a principled way by coupling NNs with Bayesian probability theory, leading to the formulation of Bayesian Neural Networks (BNNs) [39][20][16]. The most distinguishing property of a BNN is *marginalization*, i.e., rather than using a single set of the weights $\hat{\mathbf{w}}$, determined at the end of the training phase, BNNs rely on the computation of the predictive distribution for a given input $\mathbf{x}$, as follows [66]:

$$p(\boldsymbol{y}|\mathbf{x}, \mathcal{D}) = \int p(\boldsymbol{y}|\mathbf{x}, \mathbf{w}) p(\mathbf{w}|\mathcal{D}) d\mathbf{w} \tag{1}$$

where: $p(\mathbf{w}|\mathcal{D})$ is the posterior distribution over model parameters, inferred from a given data-set $\mathcal{D}$ through the Bayes theorem, starting from a prior distribution $p(\boldsymbol{w})$ [44]. Equation 1 can be also viewed as a *Bayesian Model Averaging*, where we have an ensemble of models with different parameters settings, and the overall predictions are achieved by an average over the models ensemble, weighted by their posterior probabilities [66].

The posterior distribution over the weights in Eq. 1 allows for capturing the model uncertainty, arising from the uncertainty associated with the parameters of the model, given the training dataset. In fact, it is possible to recover the usual non-Bayesian prediction $p(\boldsymbol{y}|\boldsymbol{x}, \hat{\boldsymbol{w}})$, which depends on a particular setting of the weights $\hat{\boldsymbol{w}}$, by approximating the posterior distribution over the weights with a delta distribution $p(\boldsymbol{w}|\mathcal{D}) \sim \delta(\boldsymbol{w} - \hat{\boldsymbol{w}}_{\mathrm{MAP}})$. $\delta$ is the Dirac distribution, which is zero everywhere except at the maximum of the posterior $\hat{\boldsymbol{w}}_{\mathrm{MAP}} = arg\,max_{\boldsymbol{w}} p(\boldsymbol{w}|\mathcal{D})$ [39][66].

In addition to properly quantifying the uncertainty associated with each prediction, BNNs can also offer a principled decomposition of aleatoric and epistemic uncertainty. For classification problems, it is possible to estimate the total predictive uncertainty through the *Shannon Entropy* $\mathbb{H}$ of the predictive distribution [44], which is maximized in case of a flat distribution over the classes (i.e., the most uncertain scenario). Moreover, the uncertainty of the predictive distribution can be further decomposed [16], as follows:

$$\begin{aligned} \underbrace{\mathbb{H}[p(\boldsymbol{y}|\mathbf{x}, \mathcal{D})]}_{\text{Total Uncertainty}} \;=\; & \underbrace{\mathbb{I}[\boldsymbol{y}, \boldsymbol{w} \mid \boldsymbol{x}, \mathcal{D}]}_{\text{Model Uncertainty}} \;+ \\ & \underbrace{\mathbb{E}_{p(\boldsymbol{w}|\mathcal{D})}\left[\mathbb{H}[p(\boldsymbol{y}|\mathbf{x}, \boldsymbol{w})]\right]}_{\text{Aleatoric Uncertainty}} \end{aligned} \tag{2}$$

where: $\mathbb{I}$ is the *information gain* between parameters and output, and captures the model uncertainty, while the second one is the aleatoric uncertainty, computed as the expected value $\mathbb{E}$ of the entropy of the predictions obtained by exploiting the models of the ensemble. It is crucial to decouple these two quantities since they behave differently [16]: the former is typically high for previously unseen inputs, while the latter is high for ambiguous or noisy samples and it does not decrease by acquiring more training data. In other words, the latter one cannot be exploited to enhance the quality of a model to recognize unseen inputs.

Unfortunately, the exact evaluation of the predictive distribution is computationally intractable for Neural Networks of practical size [20] [44]. To get around this problem, several approximate inference methods have been proposed. Most approaches rely on *Variational Inference* for finding a tractable approximation to the Bayesian posterior distribution of the weights [9] or on *Deep Ensembles*, where the same model is trained multiple times and then the resulting models are averaged [66].

More specifically, the Variational Inference approach aims at approximating the posterior $p(\mathbf{w}|\mathcal{D})$ with a tractable distribution $q(\mathbf{w}|\boldsymbol{\theta})$, and adjusting the parameters $\boldsymbol{\theta}$ to get the best approximation by minimizing the ELBO loss [39]: a common assumption for $q$ is a diagonal Gaussian distribution. In [9], the authors proposed a principled and backpropagation-compatible algorithm for minimizing the ELBO loss, which makes Variational Inference scalable to complex Neural Networks. In particular, for estimating the model uncertainty via Variational Inference, it is possible to compute the information gain by considering $N$ forward passes and sampling the weights from the variational distribution $q$ [9], as follows:

$$\begin{aligned} \underbrace{\mathbb{I}[\boldsymbol{y}, \boldsymbol{w}|\boldsymbol{x}, \mathcal{D}]}_{\text{Model Uncertainty}} \;=\; & \underbrace{\mathbb{H}\left[\frac{1}{N}\sum_{i=1}^{N} p(\boldsymbol{y}|\mathbf{x}, \boldsymbol{w}_i)\right]}_{\text{Total Uncertainty}} + \\ & \underbrace{-\frac{1}{N}\sum_{i=1}^{N} \mathbb{H}[p(\boldsymbol{y}|\mathbf{x}, \boldsymbol{w}_i)]}_{\text{Aleatoric Uncertainty}} \end{aligned} \tag{3}$$

where the total uncertainty is computed as the entropy of the predictive distribution (i.e., Eq.(1)) approximated as an ensemble average with respect to different parameters settings. It is possible to show [50] that a similar decomposition holds also for other kinds of classifiers, based on ensembles, like Random Forests. In that case, the total uncertainty is given by the entropy of the mean predictions and the aleatoric uncertainty by the mean entropy of the predictions of each classifier of the ensemble [50]. The rationale is that the epistemic uncertainty is high when members of the ensemble disagree, by assigning different and highly confident predictions to a given input. This is something that we exploited to build an uncertainty-aware RF model (see Section IV).

The main drawback of traditional Bayesian Neural Networks is that they typically require multiple forward passes at test time to estimate the mean predictive distribution. As a result, there has been a growing interest in developing methods for uncertainty quantification by employing deterministic single forward-pass neural networks to provide a reduced latency estimation. Among them, it is worth mentioning Deep Deterministic Uncertainty (DDU) [43], an NN-based approach that employs a feature-space density model as a proxy for the epistemic uncertainty and the entropy of the softmax outputs as a measure of the aleatoric uncertainty. We considered also DDU in our critical comparison of uncertainty-aware models.

### B. Out-of-Distribution detection

As already mentioned, most of ML models are trained based on the closed-world (or closed-set) assumption, where the test data is assumed to be drawn from the same distribution as the training data. However, when models are deployed in an open-world scenario, test samples can be Out-of-Distribution and therefore should be handled with caution, by rejecting them or handing them over to human domain experts [67].

*OoD Detection* is a broad topic, and here we concentrate only on the sub-field of *open-set classification* where a model must not only be able to distinguish between the training classes, but also indicate if an input comes from a *semantically* new class it has not yet encountered. Moreover, in our critical comparison we will consider methods for OoD Detection that do not need training or fine-tuning on OoD data, since these samples may not be available in practical applications. For a survey on OoD Detection the reader should refer to [67].

Early works observe the overconfidence of Neural Networks and therefore focus on redistributing the logits (i.e., unnormalized Softmax values). In particular, one of the first methods was OpenMax [6], which replaces the softmax layer with an OpenMax layer and calibrate the logits with a per-class probabilistic model based on the activation patterns in the penultimate layer of the Neural Network. A more recent and competitive approach in terms of OoD Detection was proposed by [38]. They suggest an Energy-Based OoD Detection method that uses a scalar energy score, which is lower for observed data and higher for unobserved ones. Moreover, this approach can be applied to any pre-trained neural classifier, without the need of re-training it or to modify its architecture. Given its advantages and peculiarities, we will consider such an Energy-Based model in our critical evaluation.

OoD Detection can be also seen as an application of epistemic uncertainty quantification: since we do not train on OoD data, we expect OoD data points to have higher epistemic uncertainty than in-Distribution (iD) data. Several works [43][60][36] explicitly leveraged this observation for effective uncertainty quantification and competitive OoD Detection.

In general, the Out-of-Distribution detection problem can be formulated as a *binary classification problem*: for a given input $\boldsymbol{x}$ and a given (close-set) classifier $f$ the following decision rule $g$ is applied:

$$g(\boldsymbol{x}, f) = \begin{cases} f(\boldsymbol{x}), & \text{if } S(\boldsymbol{x}, f) < \tau \\ \text{unknown}, & \text{if } S(\boldsymbol{x}, f) > \tau \end{cases} \quad (4)$$

where: $S$ is a score associated to a certain input for distinguishing between known and unknown inputs (e.g. epistemic uncertainty), and $\tau$ is a threshold that has to be carefully defined.

## IV. PROBLEM STATEMENT AND MODELS

In this Section, we formalize the three closely-related problems that we investigate in this paper and then we briefly describe the models used to address those problems in the context of network intrusion detection.

### A. Problem statement

We consider the following three problems, that ideally should *all* be addressed by a ML-based model adopted for network intrusion detection.

**Problem 1 (Closed-Set Classification).** The first problem deals with uncertainty quantification on a common (closed-set) multi-class classification problem. A desirable property of an algorithm employed for IDS would be to assign a high degree of uncertainty to its erroneous predictions so that a system administrator may be able to make better decisions and likely avoid severe issues. In fact, a classifier can leverage accurate uncertainty estimates to refrain from making predictions if the uncertainty degree associated to a certain fraction of samples, let's say $p$, is excessively high, and may ask for the intervention of a human expert. In such case, the classifier will only make predictions on the remaining $(1 - p)$ fraction of samples, i.e., the ones whose prediction is more certain. By relying on its capability of assessing uncertainty effectively, the classifier will prioritize making predictions on instances where it feels more confident. Taking into account two sources of uncertainty (i.e., aleatoric and epistemic) can help well assess uncertainty in closed-set scenarios.

**Problem 2 (Active Learning)**. The second problem we focus on is related to acquiring and labelling large volumes of network traffic for training a ML classifier. In the domain of IDS, the process of acquiring data and label them is complex and demanding, and there is the need of continuously acquire new labeled data [5][40][32]. In this context Active Learning [15] aims at training a ML model in a data-efficient manner by iteratively acquiring only the most relevant samples from a large pool of unlabelled data, rather than annotating all samples, and labelling them with the help of an expert. In particular, in the *active-learning loop*, everything starts by training a model on a small random batch of labeled data. Then new, informative samples are added to the original training set and the model is trained on the updated training set. This procedure is repeated until the model achieves a desirable classification performance. The ultimate goal is to make the model ensure optimal performance while using the minimum amount of training data.

Equation (2) offers a natural and principled way for measuring how a sample is informative for a given model, since the mutual information $\mathbb{I}[\boldsymbol{y}, \boldsymbol{w} \mid \boldsymbol{x}, \mathcal{D}]$ expresses how much knowing the label $\boldsymbol{y}$ of a given sample $\boldsymbol{x}$ will reduce our uncertainty about the model parameters $\boldsymbol{w}$, and thus can be used as an acquisition strategy for selecting points to label. This approach, known in the literature as Bayesian Active Learning by Disagreement (BALD) [30], has proven to be highly effective in the context of Deep Learning (e.g. [21][51]). Active Learning effectiveness can also be seen as an additional evaluation of the ability of a model to estimate the epistemic uncertainty and disentangle different sources of uncertainty [43].

**Problem 3 (OoD Detection)**. The third problem that we tackle in this paper can be summarized as follow: given a classifier $f$ trained using a dataset consisting of $K$ known classes (e.g. known kinds of attacks), is it able to recognize as *unknown*, and thus abstain from performing the classification, inputs which belongs to new, *semantically* different, classes?

While many ML-based IDS have been proposed in the literature, the vast majority of them are focused on enhancing closed-set classification performance, while not analyzing their predictive uncertainty and the OoD Detection problem, as highlighted in Section II and III. In the context of intrusion detection, truthful predictions are crucial for early identification of potentially anomalous network traffic, e.g. new kinds of attacks or variations of known attacks, allowing a network operator to proactively take risk-informed countermeasures. For instance, an alarm can be triggered if the observed uncertainty or a custom score for OoD Detection exceeds a predefined threshold set by the network operator upon experience. This threshold may be established by considering known traffic patterns and serves as basis for identifying novel potential intrusions. By comparing the score against the threshold, it is possible to identify suspicious activities that deviate from normal behavior, alerting network operators to potential security threats.

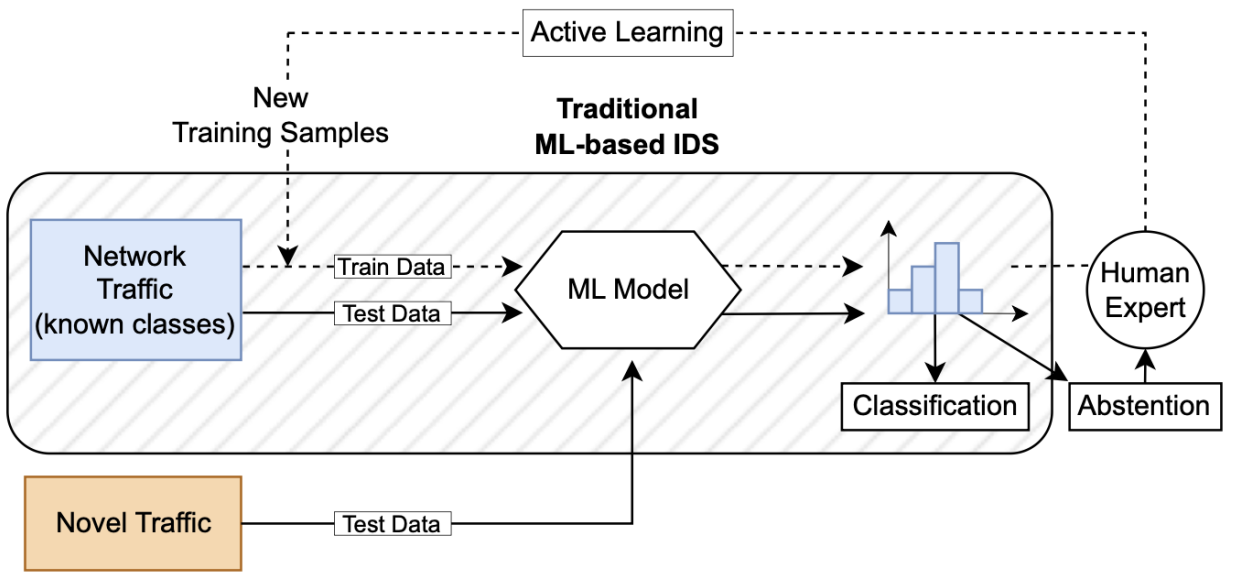

FIG. 1. Pictorial representation of the proposed approach: an uncertainty-aware ML model is able to recognize unknowns or ambiguous inputs and require the intervention of a human expert. Moreover, by estimating the epistemic uncertainty, it is possible to perform the Active Learning loop, thus training the model with the smallest possible amount of data.

The overall picture of the considered approach for developing an IDS is summarized in Fig. 1. The uncertainty-aware model is adopted by an IDS, which can be either a software artifact running in a host or in an Edge Computing node, or a hardware appliance placed at the edge. It is locally trained and is able to acquire in real time packet-related data by means of *packet mirroring*, executed by a border switch, or by exploiting efficient feature extraction capabilities of innovative *programmable data planes* of networking devices [18]. The model acts on a labeled dataset of network traffic, possibly involving the Active Learning loop, and it can provide truthful uncertainty estimates of incoming traffic, possibly asking an expert (e.g. a security engineer) for intervention: this may happen for ambiguous iD inputs or in the case of OoD samples. In the latter situation, unknowns can then be added to the training data, upon appropriate labelling by the human expert, so that the model can be retrained also on new kinds of intrusions.

Once again, we would like to stress that uncertainty-aware models are an appealing foundation for implementing the described pipeline and tackling the three main problems previously described in this Section. In the end, the ambition is to develop a model that behaves as expected across different tasks (e.g. proper uncertainty quantification and OoD detection) and ultimately increase one's trust in it.

### B. Models

In this Section, we briefly describe the main peculiarities of the models compared and evaluated in this paper. The detailed implementation of each of them is instead described in Section VI A. Table I summarizes the considered approaches and their ability to perform proper

TABLE I. Summary of the considered models and their peculiarities

| Model | NN-Based | BNN-Based | Aleatoric and Model Uncertainty Decomposition | Active Learning | OoD Detection |
|---|---|---|---|---|---|
| NN | ✓ | ✗ | ✗ | ✗ | ✗ |
| Energy-Based [38] | ✓ | ✗ | ✗ | ✗ | ✓ |
| DDU [43] | ✓ | ✗ | ✓ | ✓ | ✓ |
| BNN | ✓ | ✓ | ✓ | ✓ | ✓ |
| RF | ✗ | ✗ | ✓ | ✓ | ✓ |
| EFC [57] | ✗ | ✗ | ✗ | ✗ | ✓ |
| UC-BNN (Ours) | ✓ | ✓ | ✓ | ✓ | ✓ |

*uncertainty quantification* (by decomposing aleatoric and model uncertainty), *OoD Detection* and *Active Learning*. The symbol ✓ indicates that a model possesses the considered ability. In addition, the Table also specifies whether the approach is based on a NN model or not.

We considered four different types of NN-based models:

- **Neural Network (NN)**: This is our *uncertainty-unaware baseline* model. We consider a Multi-Layer Perceptron (MLP), which is a deterministic model able to provide the conditional probability distribution $p(\boldsymbol{y}|\boldsymbol{x}, \mathcal{D})$ over $K$ classes through the softmax activation function [44]. It is common to use the entropy of the softmax distribution as a measure of uncertainty about the classification of a given input. However, different studies [28][55] have shown that this quantity presents issues in uncertainty quantification, especially considering OoD data, due to the fact that softmax entropy is inherently not able to capture epistemic uncertainty, as previously discussed. This architecture will be exploited as a basis for other models and methods discussed in the following.

- **Energy-Based** [38]: This approach is specifically designed for OoD Detection without leveraging on uncertainty quantification. The authors propose an *energy score* to differentiate between OoD and iD samples and mitigate the critical problem of softmax's low confidence with arbitrarily high values for OoD examples. Specifically, given a trained neural network, denoting the logits corresponding to the $y$-th class label as $f_y(\boldsymbol{x})$, they define the energy score as $E(\boldsymbol{x}, f) = -T \log \sum_{i=1}^{K} e^{f_i/T}$, where $K$ is the number of classes and $T$ is a free parameter, typically equal to 1 [38]. During the test phase inputs with higher energies are considered as OoD samples and vice versa. The main advantage of this method is that it can be applied to a given NN, without the need of modifying the architecture or retrain it.

- **Deep Deterministic Uncertainty (DDU)** [43]: It is based on approximating the *feature space distribution* as a Gaussian mixture model, through a Gaussian Discriminant Analysis (GDA). More specifically, let $\boldsymbol{z}$ be the feature representation of an input $\boldsymbol{x}$. DDU involves computing the feature density $p(\boldsymbol{z})$, as a proxy for the epistemic uncertainty, by marginalizing over the classes' distributions: $p(\boldsymbol{z}) = \sum_i p(\boldsymbol{z}|c_i)p(c_i)$, where $p(\boldsymbol{z}|c_i)$ is the conditional probability distribution (modelled as a Gaussian) for each class $c_i$ , and $p(c_i)$ is the per-class prior. As a consequence, for a given input, it is possible to avoid the computation of multiple forward passes through the NN at test time, since a proxy of the epistemic uncertainty is estimated by evaluating the density of the feature representation given by the Gaussian mixture density model. The key observation of [43] is that density-based or distance-based models in the hidden layers of NNs may fail due to the feature-collapse problem [61]: feature extractors might map the features of OoD inputs to iD regions in the feature space, without a suitable inductive bias. As a consequence, DDU proposes to rely on spectral normalization [41] in models with residual connections [27], in order to encourage a distance-preserving hidden representation of the inputs and hence enhance the OoD detection.

- **Bayesian Neural Network (BNN)**: Also in this case, the architecture is the same as that of NN (e.g. same number of layers, neurons, etc.). The key difference is that in a Bayesian setting the goal is to learn from a training dataset an ensemble of plausible parameters in the form of a posterior probability density, rather than a point estimate. To get the posterior it must be specified a prior over the weights (see Section III). Here we employ the common *zero-mean, isotropic Gaussian distribution* as a prior over the weights: $p(\boldsymbol{w}) = \mathcal{N}(0, \alpha^2\boldsymbol{I})$, where the variance $\alpha^2$ was chosen to recover the weight decay $\beta$ of the *L2 regularizer* employed for the NN in the following way: $\alpha^2 = 1/2\beta$ [66][44]. Bayesian predictions involve marginalization over the parameters: to evaluate Equation 1, we leveraged the Variational Inference approach described in Section III A.

Finally, it s important to observe that BNNs may not consistently exhibit elevated uncertainty for OoD inputs, particularly in the case of simple models, like Multi-Layer Perceptron [44][66]. In this case, there might be regions of the input space where simple BNNs can exhibit overconfident predictions, even for inputs far away from the training data (i.e., OoD samples). This observation serves as the starting point for developing our custom model to address this undesired behavior.

- **Proposed approach (UC-BNN)**: Our main goal is to propose a model that shows all the benefits of a BNN model in closed-set classification settings (i.e., meaningful uncertainty estimates and the ability to perform Active Learning) and boost its capabilities of detecting OoD inputs, thus reducing false alarms and without significantly increasing the computational overhead. To achieve this we propose a method for gauging the predicted total and espistemic uncertainty of a given trained BNN model to improve the OoD detection without affecting the iD predictions. We call this model Uncertainty-Corrected BNN (UC-BNN). The core ideas behind the definition of this model are: *i)* Building a meaningful and distance-aware feature extractor, by avoiding the feature collapse phenomenon, as previously described in the NN-based DDU model [43]. *ii)* After training the BNN, fitting a density model in the feature space of the training data. *iii)* Exploiting the density model to compute the probability density function (pdf) associated to each input, and use this quantity to re-calibrate both the total and epistemic uncertainty, to reduce overconfident predictions for inputs which are far from the training set. A visual representation of the proposed approach is depicted in Fig. 2.

  To reduce the feature collapse in the feature space and improve sensitivity (step *i)*) we add a skip connection to each layer [43][37] of the previously described BNN model, so that the output of each layer is now computed as $z_{\mathrm{OUT}} = x_{\mathrm{IN}} + f(x_{\mathrm{IN}})$, where $f$ denotes the activation function, instead of the usual $z_{\mathrm{OUT}} = f(x_{\mathrm{IN}})$. Moreover, as activation functions, we employ LeakyReLU [14] in the first hidden layer and ELU [14] in the second, to further improve sensitivity, since also negative activations can be propagated trough the network, and to provide more Gaussian-like activations with respect to the ReLU activation function [14]. After implementing such a distance-aware feature extractor, we model the hidden (or latent) distributions of the training data in the penultimate layer of the network as a multivariate Gaussian (step *ii)*), parameterized by a mean $\overline{\boldsymbol{z}}$ and a covariance matrix $\Sigma$. Under this assumption, the probability density associated to the hidden representation $\boldsymbol{z}$ of a given input is uniquely determined by the so-called Mahalanobis distance [Bishop] between the mean of the feature distribution $\overline{\boldsymbol{z}}$ and $\boldsymbol{z}$ itself:

$$d(\boldsymbol{z}, \overline{\boldsymbol{z}}) = \sqrt{(\boldsymbol{z} - \overline{\boldsymbol{z}})^{\mathrm{T}} \Sigma^{-1} (\boldsymbol{z} - \overline{\boldsymbol{z}})}. \tag{5}$$

The next step involves recalibrating the uncertainty for each input (step *iii)*). This process relies on the intuition that we expect a high epistemic uncertainty for the farthest inputs (with reference to the Mahalanobis distance), or, equivalently, for those inputs whose hidden representation falls in the tails of the latent distribution of the training data. The proposed procedure can also be heuristically understood as placing a prior on the expected uncertainty predicted by a model, which should be high for inputs far away from the training data. Considering that the entropy is an additive quantity [Bishop], we propose to gauge the total and epistemic uncertainties provided by the BNN model by adding to them a contribution $\delta$ that depends on the distance $d(\boldsymbol{z}, \overline{\boldsymbol{z}})$ as follow:

$$\delta(d) = \begin{cases} 0 & \text{if } d \le \tilde{d} \\ 2\epsilon(\mathrm{sigmoid}(\frac{d}{\tilde{d}} - 1) - \frac{1}{2}) & \text{if } d > \tilde{d} \end{cases} \tag{6}$$

where: $\epsilon$ is the maximum entropy achivable by a flat distribution over $N$ (known) classes and $\tilde{d}$ is a threshold over the distances in the features space. We set this threshold as the 99.5% percentile of the distances calculated on the validation set, with the goal of not affecting the predictions on iD data and mitigating the influence of potential outlier values. The particular functional form of Eq. (6) was chosen to be a continuous function, which essentially does not correct the predictions over the iD data and smoothly increases the uncertainty prediction far away from the training data, up to the maximum achievable entropy $\epsilon$. The overall procedure is summarized by the pseudocode in the Algorithm 1. It should be noted that the proposed correction applies to both the total and epistemic uncertainty (lines 9 and 12 of Algorithm 1) so that the aleatoric uncertainty is unchanged, as well as the uncertainty associated to predictions that are already maximally uncertain. In fact, lines 10-12 ensure that both the total entropy and the epistemic uncertainty are at most equal to the maximum entropy $\epsilon$, when $\mathbf{T} + \delta > \epsilon$. In this case, the calibration is determined by the difference between the maximum entropy and the total uncertainty predicted by the model.

Last, we emphasize that the recalibration procedure applies after the training of the model, which is when the hidden feature distribution of the training data can be considered fairly representative of the iD data distribution. In the Active Learning process this assumption might not be valid, especially during the first iterations. For this reason, UC-BNN relies on the epistemic uncertainty directly offered by the BNN model to perform Active Learning.

**Algorithm1** Uncertainty correction (UC-BNN)

**Require:** Trained BNN
**Require:** Mean feature vector $\overline{z}$ and covariance matrix $\Sigma$
**Require:** Threshold $\tilde{d}$
**function** CALIBRATION FUNCTION(input $\boldsymbol{x}$):
  Let $\boldsymbol{z}$ be the hidden representation of an input $\boldsymbol{x}$
  Let $\epsilon$ be the maximum entropy achivable
  Predict the class-probabilities $\boldsymbol{y}$ of $\boldsymbol{x}$
  Compute Epistemic $\mathbf{E}$ and Total $\mathbf{T}$ uncertainty
  Compute $d(\boldsymbol{z}, \overline{z})$ ▷ as in Eq.(5)
  Compute correction $\delta(d)$ ▷ as in Eq.(6)
  **if** $\mathbf{T} + \delta \leq \epsilon$ **then**
    **return** $\mathbf{T} + \delta$, $\mathbf{E} + \delta$
  **else**
    gap $= \epsilon - \mathbf{T}$
    **return** $\mathbf{T} + \text{gap}, \mathbf{E} + \text{gap}$
  **end if**
**end function**

In addition, we considered two different promising models, not based on NNs:

- **Random Forest (RF)**: It is a classification method based on an ensemble of decision trees where each member of the ensemble is trained with bootstrapped samples of a given training set. Each tree can predict the class probability for a given input, which is estimated as the fraction of samples of the same class in a leaf. Then, the overall prediction of the forest is achieved by averaging the predicted class probabilities over different trees, in contrast to the usual approach of majoring vote. Since this model is an ensemble of different predictors, following [50] it is possible to exploit the same decomposition of total uncertainty reported in Equation 3 (in this case the sum runs over the number of trees of the forest), to estimate the aleatoric and epistemic uncertainty, and to make RF inherently uncertainty-aware.

- **Multi-class Energy-based Flow Classifier (EFC)** [57]: This model is specifically designed for OoD Detection in an intrusion detection scenario without leveraging on uncertainty quantification: it is a statistical model designed specifically for classifying network flows. With EFC multiple distributions are inferred, with each distribution representing a distinct flow class. Subsequently, flow energies are computed within each distribution, and these values are compared to determine the final classification outcome (i.e., known class or unknown).

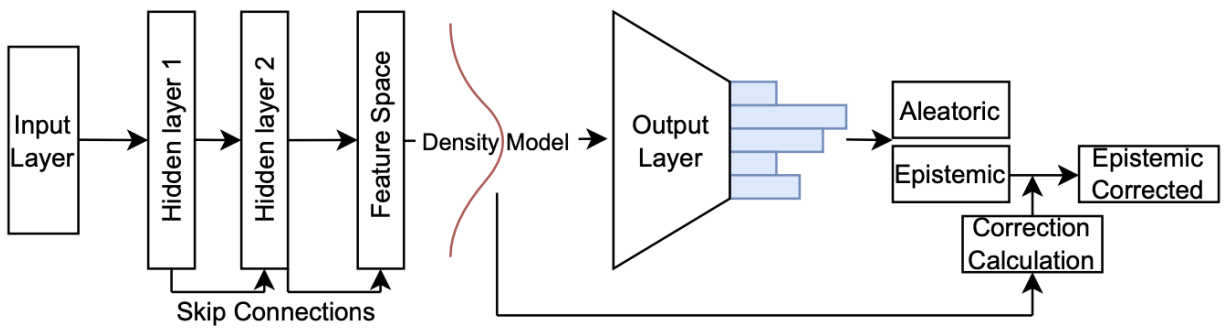

FIG. 2. Pictorial representation of the proposed model showing the architectural changes with respect to a state-of-the-art BNN and the recalibration process of the uncertainty.

## V. DATASET DESCRIPTION AND PREPROCESSING

We consider the open-source dataset "NF-ToN-IoT-v2" [2] which includes several network-related attacks related to Internet-of-Things (IoT) devices [49]. This is a realistic dataset based on a IoT testbed composed of edge computing nodes and a cloud infrastructure to simulate an IoT production environment [42], and it is often used in the literature as a benchmark (e.g. [11]). The dataset includes real collected data and is based on 43 NetFlow [13] features extracted from the network packets (e.g. L4 source port, L4 destination port, TCP flags, etc.), describing the network traffic between different sources and destinations identified by their IP addresses [49]. It is organized *per-flow*: each row represents a flow (i.e., source/destination IP pair), for which some additional statistics are computed and also considered as features (e.g. longest flow packet, shortest flow packet, flow duration, etc.) and is annotated as belonging to one over a total of 10 classes, including benign traffic and nine different attacks. Among the overall features, we removed from the dataset the timestamp and the source/destination IP addresses in order to only look for intrinsic patterns in the network traffic flows.

Table II summarizes the nine types of attacks, as long as the distribution of samples (i.e., flows) for each class. In general, we consider two testing scenarios: the usual *closed-set classification* (exploited also for *Active Learning* experiments) and *OoD Detection.* To simulate OoD inputs appearing at run time, we entirely remove multiple classes from the original dataset and consider them as unknowns. We treat the remaining samples, from the remaining known classes, as a training set to develop a usual multi-class classifier. The dataset with the known classes is partitioned into 60% training, 20%, validation, and 20% testing set, and standardized using the training data of known classes, so that each feature distribution presents a zero mean and unit variance. Also the unknowns are standardized by exploiting the features means and variances computed on the training data. Only with the Random Forest classifier [50] we did not standardized the data, since it is not required.

[2] Accessible at https://staff.itee.uq.edu.au/marius/NIDS_datasets/#RA7

TABLE II. Samples distribution

| Class | Number of samples | Scenario | | |
|---|---|---|---|---|
| | | 3U | 6U | 8U |
| Benign | 6099469 | K | K | K |
| Scanning | 3781419 | K | K | K |
| XSS | 2455020 | K | K | U |
| DDoS | 2026234 | K | K | U |
| Password | 1153323 | K | U | U |
| DoS | 712609 | K | U | U |
| Injection | 684465 | K | U | U |
| Backdoor | 16809 | U | U | U |
| MITM | 7723 | U | U | U |
| Ransomware | 3425 | U | U | U |

We consider different partitions of the dataset by varying the number of classes moved in the OoD dataset (i.e., whose samples are unknowns) to test the models with different distributions of knowns/unknowns. In particular, we considered as knowns the classes with more samples, in order to mimic a realistic case where an IDS is trained on the well-known and most representative types of network traffic (i.e., the benign traffic and the most common attacks). The considered scenarios are called 3U, 6U and 8U in Table II, where the number of the abbreviation indicates how many and what classes are unknowns (U) and knowns (K).

As a reference setup for the closed-set classification and Active Learning experiments we considered the 3U scenario. For further analysis on the OoD Detection setting, we then varied the number of known (K) and unknown (U) classes to check the robustness of the models with respect to changes in the known/unknown distributions. It should be specified that in all the cases we randomly sub-sampled the unknown classes to obtain the same number of per-class unknowns. The reason is that in this way we test models treating the possible unknown distribution equally, without favoring any particular region of the input space.

## VI. NUMERICAL RESULTS

### A. Models implementation details

Before getting into the numerical results, we report some details on how we implemented the considered models:

- **NN**: We considered a Neural Network with 2 fully connected hidden layers with 64 neurons each and ReLU as activation function [44].To reduce overfitting we added 2 layers of batch normalization [31] after each single hidden layer. Moreover, for each layer, we employed the L2 regularizer with a weight decay $\beta = 0.1$ [44]. The training process is performed using the Adam optimizer [34] with an initial learning rate of $10^{-2}$ and the cross-entropy as loss function [44]. The training data is presented to the NN in batches of 128 samples for up to 10 epochs. All the parameters and hyperparameters were chosen to minimize the loss on the validation set: we explored different settings leveraging the Tree Parzen Estimator (TPE) Bayesian optimization algorithm implemented in the Optuna library [2]. We implemented the classifier using Google's TensorFlow platform [1].

- **Energy-Based** [36]: We considered the same implementation of NN with the only difference that we added, in the form of a Python script, the energy-score computation from the logits.

- **DDU** [43]: Starting from the previously-described NN architecture, we added in the form of a Python script a residual connection to each layer and we estimated the per-class density distribution in the last hidden layer of 64 neurons, following [43]. For training this model we selected a batch size of 256 samples, which provides a more stable training on the validation set.

- **BNN**: The architecture is the same as the previous NN but without the batch normalization layers. In order to implement the Variational Inference approximation we relied on the TensorflowProbability library [17] exploiting DenseFlipout layers [65]. We trained the network for 10 epochs with batches of 128 samples.

- **UC-BNN**: The implementation follows the previous BNN model. The only difference, according to the description of the approach reported in Section IV B, is that here we added a first linear layer necessary for implementing the skip connections and we change the activations function for the two hidden layer from ReLU to LeakyReLU and ELU, respectively. Moreover, we added a skip connection to each layer and we trained the network for 10 epochs with batches of 128 samples. Lastly, the mean feature vector $\bar{z}$ and covariance matrix $\Sigma$ are estimated through the Minimum Covariance Determinant covariance estimator of [48].

- **RF**: We used the Random Forest Classifier from scikit-learn [48]. The number of trees within the forest is set to 25, while the other hyperparameters are standard from [48], and we use bootstrapping to induce diversity between the trees of the forest. The described setup has been found to maximize the validation accuracy at a reasonable computational cost.

- **EFC** [57]: All the experiments related to this model are based on the public repository made available by the authors considering its standard parameters [57].

TABLE III. Test set performance metrics

| Closed-set classification | | | | | |
|---|---|---|---|---|---|
| Model | Accuracy | F1-Weighted | F1-Macro | ECE | MCE |
| NN | $91.8 \pm 0.6$ | $91.7 \pm 0.6$ | $86.1 \pm 0.7$ | $0.777 \pm 0.006$ | $0.839 \pm 0.002$ |
| Energy-Based [36] | $91.5 \pm 0.6$ | $91.45 \pm 0.6$ | $86.5 \pm 0.7$ | $0.775 \pm 0.006$ | $0.838 \pm 0.002$ |
| DDU [43] | $92.6 \pm 0.4$ | $92.4 \pm 0.3$ | $88.0 \pm 0.6$ | $0.785 \pm 0.006$ | $0.842 \pm 0.002$ |
| BNN | $96.55 \pm 0.01$ | $96.52 \pm 0.01$ | $93.85 \pm 0.01$ | $0.810 \pm 0.002$ | $0.841 \pm 0.002$ |
| RF | $98.291 \pm 0.001$ | $98.314 \pm 0.001$ | $96.652 \pm 0.002$ | $0.817 \pm 0.0001$ | $0.854 \pm 0.001$ |
| EFC [57] | 85.11 | 86.1 | 72.3 | - | - |
| UC-BNN (Ours) | $96.50 \pm 0.01$ | $96.48 \pm 0.01$ | $93.52 \pm 0.01$ | $0.809 \pm 0.002$ | $0.839 \pm 0.002$ |

All the experiments were performed on a workstation with 64 GB of RAM, equipped with an Intel Core Xeon 8-Core CPU.

### B. Closed-Set Classification

In this Section we compare the models through the lens of the usual closed-set classification for the 3U scenario (see Table II) by considering the overall *Accuracy*, the macro F1-Score (*F1-Macro*, i.e., an arithmetic mean of the F1-Score of each class) and the weighted F1-Score (*F1-Weighted*, i.e., a mean of the F1-Score of each class weighted by the number of per-class samples). Moreover, we reported also two common metrics for measuring the *calibration* of a classifier, i.e., the Expected Calibration Error (*ECE*), and the Maximum Calibration Error (*MCE*) [23][47].

A model is said to be calibrated if its predicted probabilities match the empirical frequencies: for instance, if a classifier predicts $p(y = k \mid \boldsymbol{x}) = 0.8$ for a certain set of inputs, then we expect the class $k$ to be the true label of the related inputs for about 80% of the time. It is possible to assess calibration by dividing the predicted probabilities into a finite set of $M$ bins and then, for each bin, assess the discrepancy between confidence and accuracy of samples whose prediction confidence falls into the considered bin. More precisely, the Expected Calibration Error is defined as $\text{ECE} = \sum_{i=1}^{M} \frac{B_i}{n} |\text{acc}(B_i) - \text{conf}(B_i)|$, where $n$ is the number of samples and $B_i$ is the set of indices of samples whose prediction confidence falls into the $i$-th bin [23]. For risk-sensitive applications, where reliable confidence measures are necessary, it is possible to rely on the MCE, computed as the maximum discrepancy between confidence and accuracy over the bins [23]. The lower ECE and MCE, the better.

In Table III we report the mean value for each metric, along with its standard error. These quantities are computed by repeating the train-test loop 16 times with different random initialization of the training algorithm, to check the robustness of the classifiers with respect to the training procedure. ECE and MCE are computed by partitioning the prediction interval into 10 bins, as in [23].

Random Forest achieves the highest performance, while NN-based approaches (i.e., NN, Energy-Based, DDU, BNN and UC-BNN) are less competitive. However, it is interesting to note that BNN-based models (i.e., BNN and UC-BNN) perform significantly better than the standard NN (and related methods), despite they basically share the same architecture. The reason for that lies in *how* the two approaches (NN-based and BNN-based) make their prediction and in particular in the Bayesian model averaging of Equation 1, which is beneficial also for improving classification performance (in terms of Accuracy and F1-Score), and not only for uncertainty quantification [66][53]. Moreover, it should be noted that BBN-based models, present a more stable performance than traditional NNs, represented by a significantly smaller standard error on the different classification metrics. Note that for EFC we do not report any standard error since this quantity is negligible for this model. Finally, it is worth mentioning that NN, Energy-Based and DDU achieve comparable performance, since the first two models share exactly the same architecture, while the last one only adds residual connections and spectral normalization. Aditionally, it can be concluded that the architectural changes introduced by the proposed approach, UC-BNN, have only a marginal impact on the performance of BNN in the usual classification setting. Nevertheless, the proposed model continues to exhibit superior performance compared to NN-based models.

From the point of view of the calibration, all models are well-calibrated (ECE and MCE $< 1$) [23], so there is no need of applying a post-training calibration process. We did not reported ECE and MCE for EFC as it does not predict a probability distribution as output.

In order to further investigate how well models provide truthful uncertainty estimation on their prediction, we select a subset of predictions whose confidence is above a varying threshold and, by applying the decision rule in Equation 3, we compute the Accuracy only on the most certain outputs, rejecting the others. By increasing the threshold we expect an increase in the number of Rejections as well as an increase in the overall Accuracy. In Figure 3 we report the plot of the Accuracy-Rejections curve for non-Bayesian NN-based models (i.e., NN, Energy-Based and DDU, which perform the same), BNN models (including our proposal UC-BNN) and Random Forest. EFC cannot be evaluated from this point of view since it does not provide uncertainty on its predic-tions. More specifically, we report the mean accuracy and the standard deviation computed over 16 experiments.

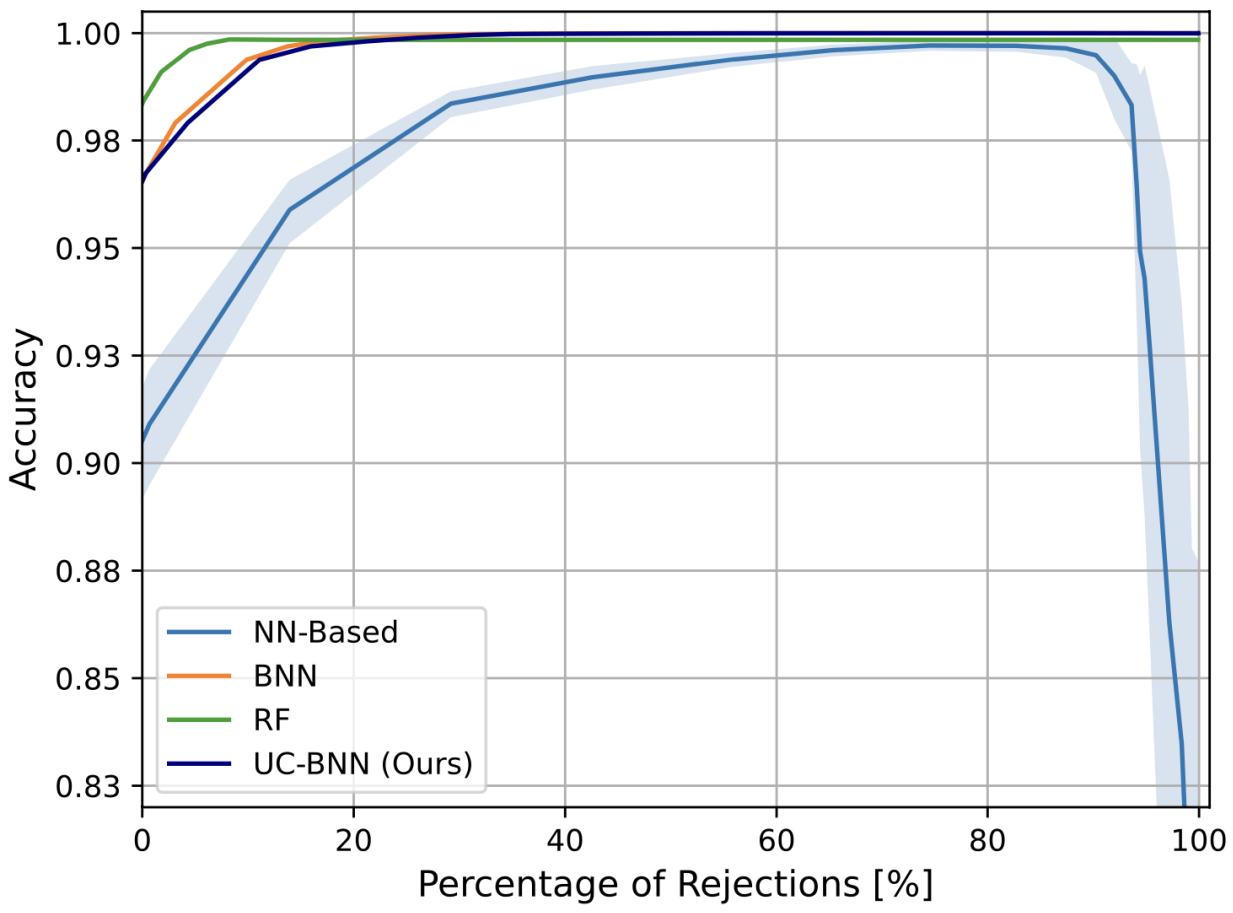

FIG. 3. Accuracy-Rejections plot for different models evaluated on the test set. The accuracy is computed only on a subset of test samples, by rejecting instances classified with a decreasing degree of uncertainty. Solid curves represent the mean accuracy, while bands represent one standard deviation (negligible for RF,BNN and UC-BNN) over 16 repeated experiments. Non-Bayesian NN-based models show overconfident predictions by wrongly classifying the most certain samples.

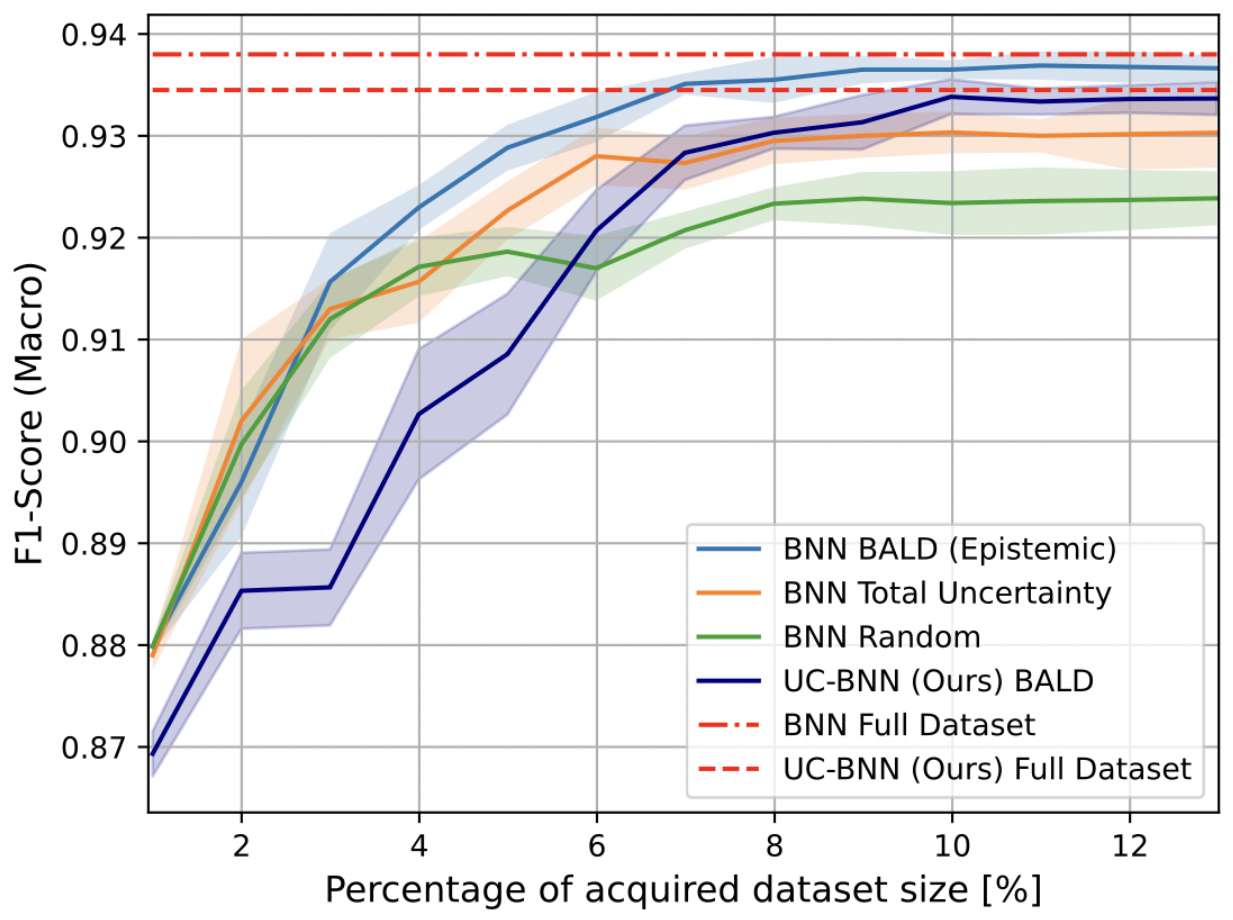

FIG. 4. Active Learning experiments with BNN-based models. Solid curves represent the mean Macro F1-Score for each acquired batch of data, while bands represent one standard deviation.

It is possible to notice that in the range of low Rejections the Accuracy increases monotonically for all models, but for high Rejection percentages (i.e., by keeping only the most certain samples), the Accuracy of non-Bayesian NN-based models drops. This reflects the fact that standard NNs tend to give misleading confidence predictions, since they assign high confidence to samples that are, in fact, wrongly classified. On the other hand, BNN-based models, including UC-BNN, show a monotonic increase in accuracy, which reflects the expected behavior of proper and meaningful uncertainty estimates, and the same phenomenon is experienced also by RF. We also found a similar behaviour for the F1-Scores, but we do not report the plot for the sake of conciseness. It is also worth noting that the proposed approach, UC-BNN, does not alter the performance of BNN in terms of uncertainty estimates on iD data, as expected during the design of this method.

In conclusion, this analysis emphasize the importance of having proper uncertainty estimates: if the confidence estimates are well-calibrated, one can trust the model's predictions when the reported uncertainty is low and rely on a different strategy (e.g. the intervention of a human expert) when the model is not confident, especially if the cost of a wrong prediction may be severe.

### C. Active Learning

As we already mentioned, the real-time collection and labelling of large volumes of network traffic present several issues. Active Learning can exploit the epistemic uncertainty estimation of a given model to acquire (and hence label) the smallest amount of data that is relevant for the training process. We evaluate different models exploiting different acquisition functions: *BALD* (i.e., the epistemic uncertainty), the *total uncertainty* (exploited for instance in [51]), and a *random* acquisition function as a baseline.

In particular, we started with $10^5$ samples (i.e., $\sim 1\%$ of the overall dataset) and added samples until the model reached the desired classification score, with an acquisition size of $10^5$ samples at each step. The acquisition size was chosen to have a reasonable trade-off between collected size and computational cost while performing the experiments. At each step, the model is trained on the test set and evaluated in terms of the Macro F1-Score, to give the same weight to each class, and these experiments have been repeated 5 times for each model and acquisition function. To summarize the results we report the mean Macro F1-Score and the standard deviation, computed on the test set, as a function of the acquired training dataset size, expressed as a percentage of the full training set.

Figure 4 shows the results for the BNN-based models, including UC-BNN: BALD acquisition function performs better than acquiring data at random and selecting points exploiting only the total predictive uncertainty. By exploiting BALD as an Active Learning acquisition strategy it is possible to reach the expected F1-Score (i.e., that obtained by considering the full training dataset) with just $\sim 11\%$ of the samples, which may be seen as a significant reduction in the effort of storing data, especially in storage and memory-constrained scenarios (e.g. Edge Computing), and in human intervention for labelling them. The results once again confirm the significant benefits brought by a decomposition of aleatoric and epistemic uncertainty. Finally, it is noticeable that the architectural differences between UC-BNN and BNN initially affect performance within the active learning loop. However, after a few iterations, even UC-BNN is capable of converging to its optimal performance by leveraging the model's predicted epistemic uncertainty, without any correction, as a sampling strategy. We can thus conclude that, in the long-term, UC-BNN does not significantly affect Active Learning capabilities of BNN despite its architectural changes, although the F1-score stabilized to a slightly smaller value.

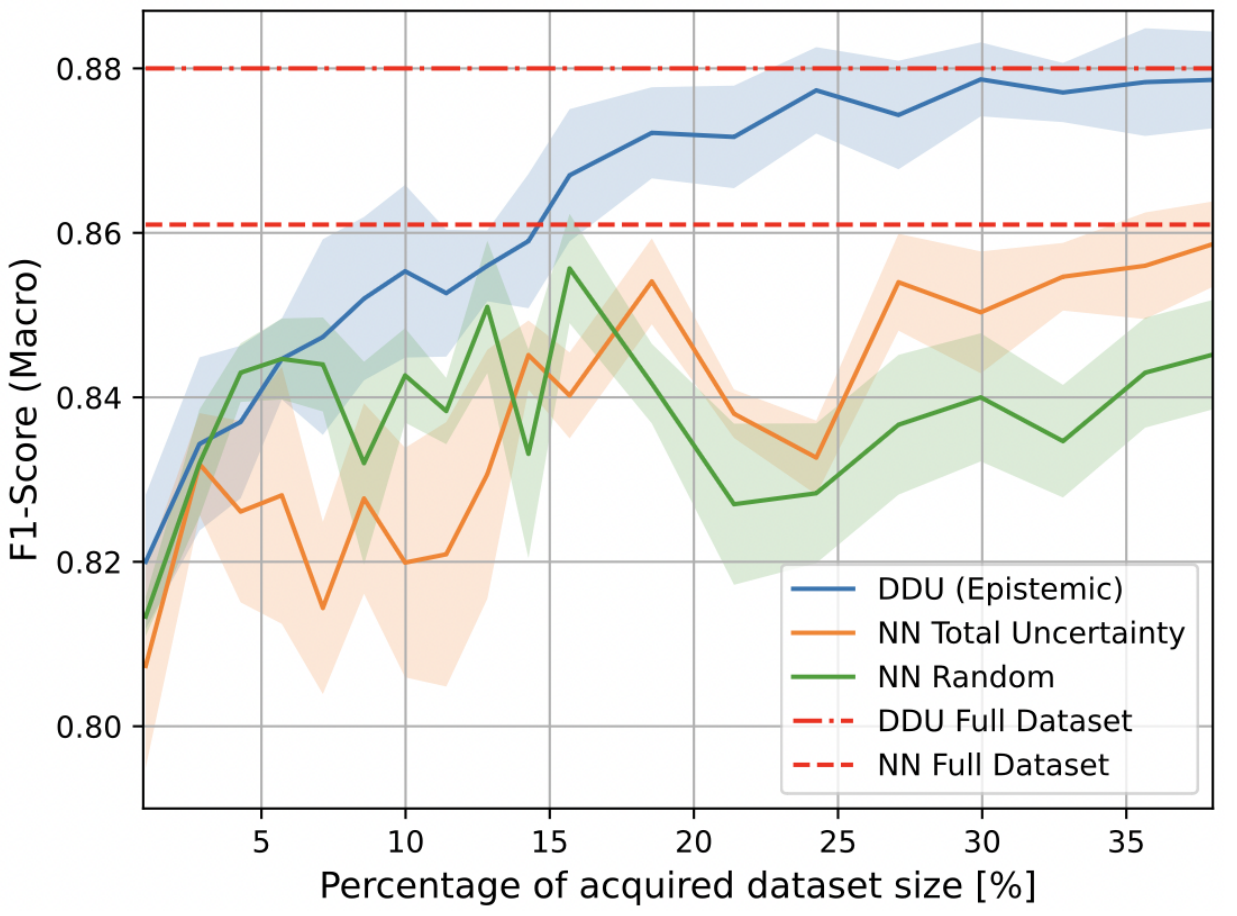

FIG. 5. Active Learning experiments with non-Bayesian NN models (i.e., standard NN and DDU). Solid curves represent the mean Macro F1-Score for each acquired batch of data, while bands represent one standard deviation.

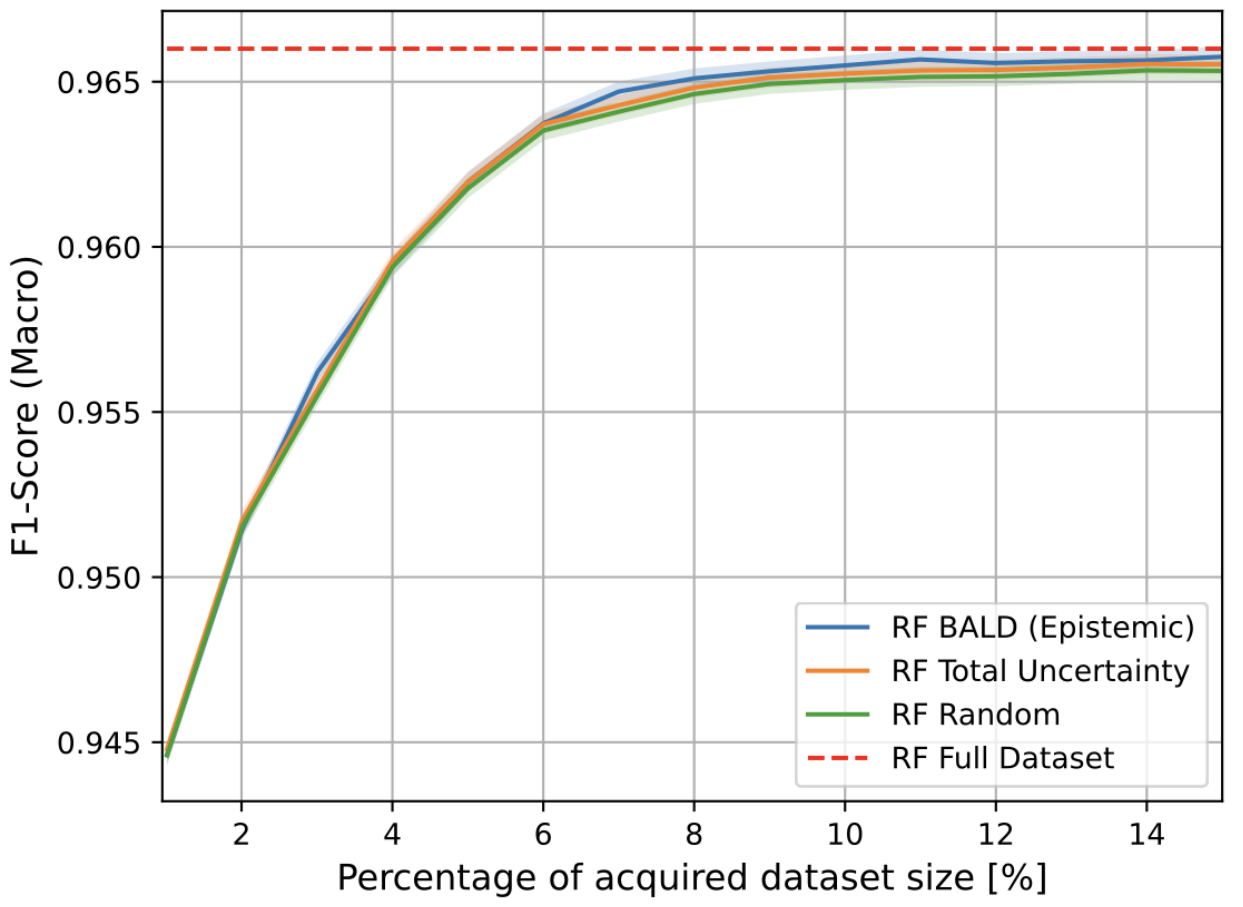

FIG. 6. Active Learning experiments with Random Forest. Solid curves represent the mean Macro F1-Score for each acquired batch of data, while bands represent one standard deviation.

Figure 5 shows the Active Learning experiments results for non-Bayesian NN-Based models. In particular, we tested NN by acquiring data exploiting a random acquisition function and the total uncertainty, given by the entropy of the softmax layers output. Energy-based is not considered as its obtained results are the same as for NN. Then, we employed DDU to sample point according to their density in the features space, so that a decomposition of epistemic and aleatoric uncertainty can be performed and the former can be used by the Active Learning loop. In this case, it was possible to obtain a faster convergence to the performance obtained with the overall dataset. However, it should be noted that traditional NN-based models (including DDU) require much more data than BNN-based ones ($\sim 35\%$ of the dataset samples) to reach the desired performance, and thus are in general less competitive for Active Learning.

Figure 6 shows the experiments for the Random Forest. In this case sampling with BALD and the total uncertainty gives almost comparable performance with respect to a random acquisition function. The fact that batch-based Active Learning does not significantly improve Active Learning for Random Forest was already noted in the literature in different contexts [46][54], not related to network traffic classification. However, the obtained results make it possible to conclude that Random Forest requires a much smaller amount of training data with respect to NN-based approaches, as with a random acquisition it is possible to reach the desired performance with only $\sim 10\%$ of samples of the full training dataset.

## D. Out-of-Distribution Detection

In this Section we discuss the numerical results for OoD Detection. In order to evaluate the performance of different models, we formulate OoD Detection as a *binary classification* problem (known vs. unknown) so that the comparison may be performed in terms of *ROC Curve* and Area Under the ROC (*AUROC*) [10] for each model (the higher AUROC, the better). In this way it is possible to fairly compare the classifiers in a threshold-independent manner and to evaluate their discriminative performance in terms of OoD Detection. Moreover, AUROC has the advantage of being insensitive with respect to the imbalance of the classes (i.e., knowns/unknowns) [10], which makes this metric particularly well-suited in our scenario.

It should be emphasized that a good IDS should operate in a regime of low False Positive Rate (*FPR*), i.e., it should provide a low rate of false unknowns. For this reason, we report also the AUROC in the region below 20% of false positives, to compare the different methods in a realistic region of deployment and assuming that 20% of false positives is the maximum acceptable value. The AUROC below the 20% of FPR has been normalized (denoted with *) so that a perfect classifier with $TPR = 1$ for each value of FPR will have an *AUROC20** of 1. Moreover, as a reference, we report also the AUROC for a random classifier.

In the first scenario (*3U scenario*), we considered the models trained on 7 known classes whose closed-set classification performance is summarized in Table III. After such training, we fed the models with input belonging to the three remaining (novel) classes. In Table IV, the results are summarized with mean values and corresponding standard errors. By looking at AUROC20*, in this setting RF seems to be the most promising method, followed by EFC, and UC-BNN among the NN-based models. Figure 7 shows the ROC curves for the considered models. It can be observed that certain models exhibit higher ROC curves in specific regions of False Positive Rate (FPR) and lower curves in other regions (e.g., DDU performs better at low FPR than BNN, but worse at high FPR). As a result, this behaviour may lead to AUROC values that are misleading when compared, and justifies our decision to focus on AUROC at low FPR (AUROC20*) for a fair comparison between methods. It is possible to notice that UC-BNN significantly improves upon BNN, especially at low false positives.

TABLE IV. Test set performance metrics with three unknown classes

| OoD Detection 3U | | |
|---|---|---|
| Model | AUROC20* | AUROC |
| NN | $0.34 \pm 0.02$ | $0.75 \pm 0.01$ |
| Energy-Based [36] | $0.38 \pm 0.02$ | $0.76 \pm 0.02$ |
| DDU [43] | $0.65 \pm 0.01$ | $0.805 \pm 0.007$ |
| BNN | $0.61 \pm 0.02$ | $0.84 \pm 0.01$ |
| RF | $0.789 \pm 0.001$ | $0.898 \pm 0.001$ |
| EFC [57] | 0.74 | 0.94 |
| UC-BNN (Ours) | $0.69 \pm 0.02$ | $0.91 \pm 0.02$ |
| Random | 0.1 | 0.5 |

To further investigate the behaviour of those models, we considered two additional scenarios by reducing the number of known classes and increasing the unknowns.

We started by considering four classes of knowns (i.e., Benign, DDoS, Scanning and XSS) and six classes of unknowns (*6U scenario*). Since here we are interested in OoD Detection, we did not reported the closed-set classification performance. However, it is worth mentioning that the closed-set performance increased by almost 2% of F1 Score and Accuracy for each classifier, with respect to the 3U scenario. Table V summarizes the metric values for novelty detection: in this scenario, UC-BNN exhibits the highest values for AUROC and AUROC20*, while BNN performs slightly better than Random Forest and DDU. For the sake of conciseness, we did not reported the plot of ROC curve for this intermediate scenario.

The last experiments were performed on the *8U Scenario*, which is represented by only two known classes (i.e., Benign and Scanning) and 8 unknown classes. This is an extreme case, where malicious traffic is mainly unknown. In this setting, the three best models are represented by RF, UC-BNN and BNN, while EFC performance drops (see Table VI). Figure 8 shows the ROC curves for the models in this scenario. In this scenario it is possible to notice that UC-BNN outperforms other NN-based models, and it exhibits the highest True Positive Rate across a wide region of interest of low False Positive Rate, between

TABLE V. Test set performance metrics with six unknown classes

| OoD Detection 6U | | |
|---|---|---|
| Model | AUROC20* | AUROC |
| NN | $0.39 \pm 0.02$ | $0.77 \pm 0.02$ |
| Energy-Based [36] | $0.50 \pm 0.02$ | $0.815 \pm 0.02$ |
| DDU [43] | $0.63 \pm 0.01$ | $0.85 \pm 0.01$ |
| BNN | $0.65 \pm 0.02$ | $0.89 \pm 0.04$ |
| RF | $0.633 \pm 0.001$ | $0.804 \pm 0.001$ |
| EFC [57] | 0.58 | 0.88 |
| UC-BNN (Ours) | $0.72 \pm 0.02$ | $0.92 \pm 0.02$ |
| Random | 0.1 | 0.5 |

TABLE VI. Test set performance metrics with eight unknown classes

| OoD Detection 8U | | |
|---|---|---|
| Model | AUROC20* | AUROC |
| NN | $0.40 \pm 0.02$ | $0.75 \pm 0.02$ |
| Energy-Based [36] | $0.51 \pm 0.02$ | $0.84 \pm 0.02$ |
| DDU [43] | $0.56 \pm 0.01$ | $0.89 \pm 0.01$ |
| BNN | $0.72 \pm 0.02$ | $0.91 \pm 0.02$ |
| RF | $0.780 \pm 0.002$ | $0.797 \pm 0.002$ |
| EFC [57] | 0.23 | 0.74 |
| UC-BNN (Ours) | $0.74 \pm 0.02$ | $0.94 \pm 0.02$ |
| Random | 0.1 | 0.5 |

$\sim 0.07$ and $\sim 0.4$

Table VII report the average results as long as the standard deviation (*STD*) over the three different scenarios. In general, it is possible to state that some methods (i.e., EFC and RF) are very sensitive to the particular knowns-unknowns setting: this is reflected on high values of the standard deviation of the AUROC, especially the AUROC20*. Moreover, it is possible to notice that RF, BNN-based models and DDU, i.e., the uncertainty-aware methods able to decouple epistemic and aleatoric uncertainty, lead to decent results in terms of OoD Detection capabilities and outperform the other approaches, including the ad-hoc ones such as Energy-Based and EFC. In particular, the best-performing models in terms of AUROC20* are RF and UC-BNN, however UC-BNN exhibits more stable and consistent performance across different scenarios, as indicated by its low value of standard deviation (STD), more than three times lower than RF. Moreover, it is possible to notice that regarding the overall AUROC metric, UC-BNN presents the highest value. Overall, it is possible to state that uncertainty-aware models are particularly interesting, since they give also good results in the usual closed-set classification and when adopted in the Active Learning loop, being thus the most promising methods to address *all* the three problems related to intrusion detection introduced in Section IV.

Overall, it is possible to notice that *uncertainty-aware RF* [50] and UC-BNN should be considered as a strong baseline in the field of network intrusion detection, al-though RF presents a higher variance in the results with respect to BNN and especially UC-BNN. Furthermore, we want to highlight that the UC-BNN method demonstrates superior performance compared to other NN-based models. This underscores the effectiveness of the proposed approach in enhancing OoD detection capabilities without adversely affecting closed-set classification performance. Last, it should be noted that standard NNs give poor performance in OoD Detection: this highlights the importance of enhancing NNs with uncertainty awareness (e.g. by adopting BNN-based models or DDU) for developing a NN-based trustworthy IDS.

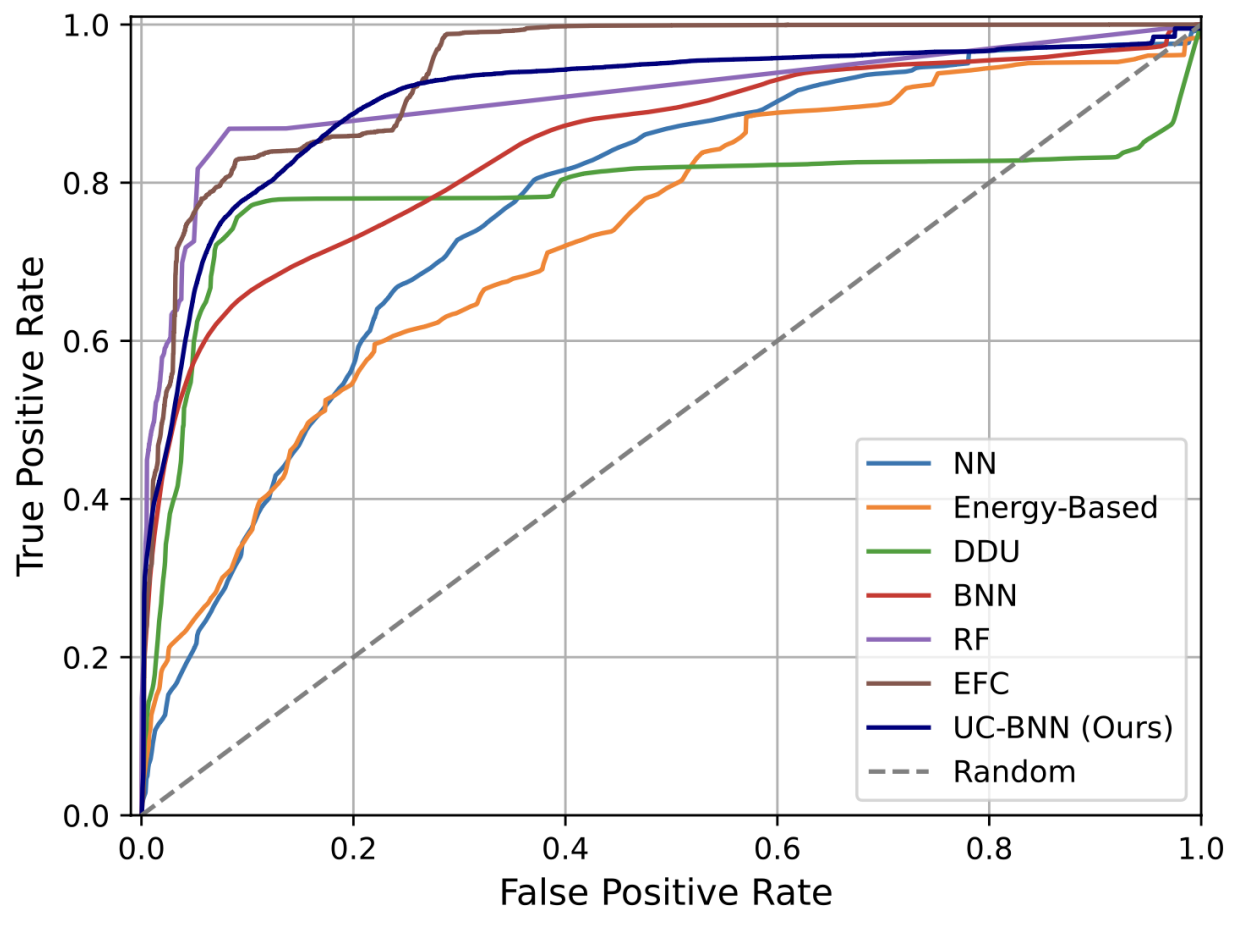

FIG. 7. Mean ROCs for the 3U scenario.

TABLE VII. Overall test set performance metrics

| OoD Detection (Average) | | | | |
|---|---|---|---|---|
| Model | AUROC20* | | AUROC | |
| | Mean | STD | Mean | STD |
| NN | 0.38 | 0.03 | 0.76 | 0.01 |
| Energy-Based [36] | 0.46 | 0.06 | 0.80 | 0.03 |
| DDU [43] | 0.63 | 0.05 | 0.85 | 0.03 |
| BNN | 0.66 | 0.05 | 0.88 | 0.03 |
| RF | 0.73 | 0.07 | 0.83 | 0.05 |
| EFC [57] | 0.55 | 0.23 | 0.85 | 0.08 |
| UC-BNN (Ours) | 0.72 | 0.02 | 0.92 | 0.014 |

## VII. CONCLUSION

This paper focuses on three fundamental and closely-related problems in the field of trustworthy ML-based network intrusion detection: *(i)* avoiding dangerous overconfident predictions in typical closed-set classification settings, *(ii)* performing efficient Active Learning on incoming network flows, and *(iii)* enabling Out-of-Distribution Detection to effectively identify unknowns. We argue that these problems should *all* be addressed, and the goal of our research has been assessing various ML models to solve them, with particular interest on methods able to guarantee proper uncertainty quantification.

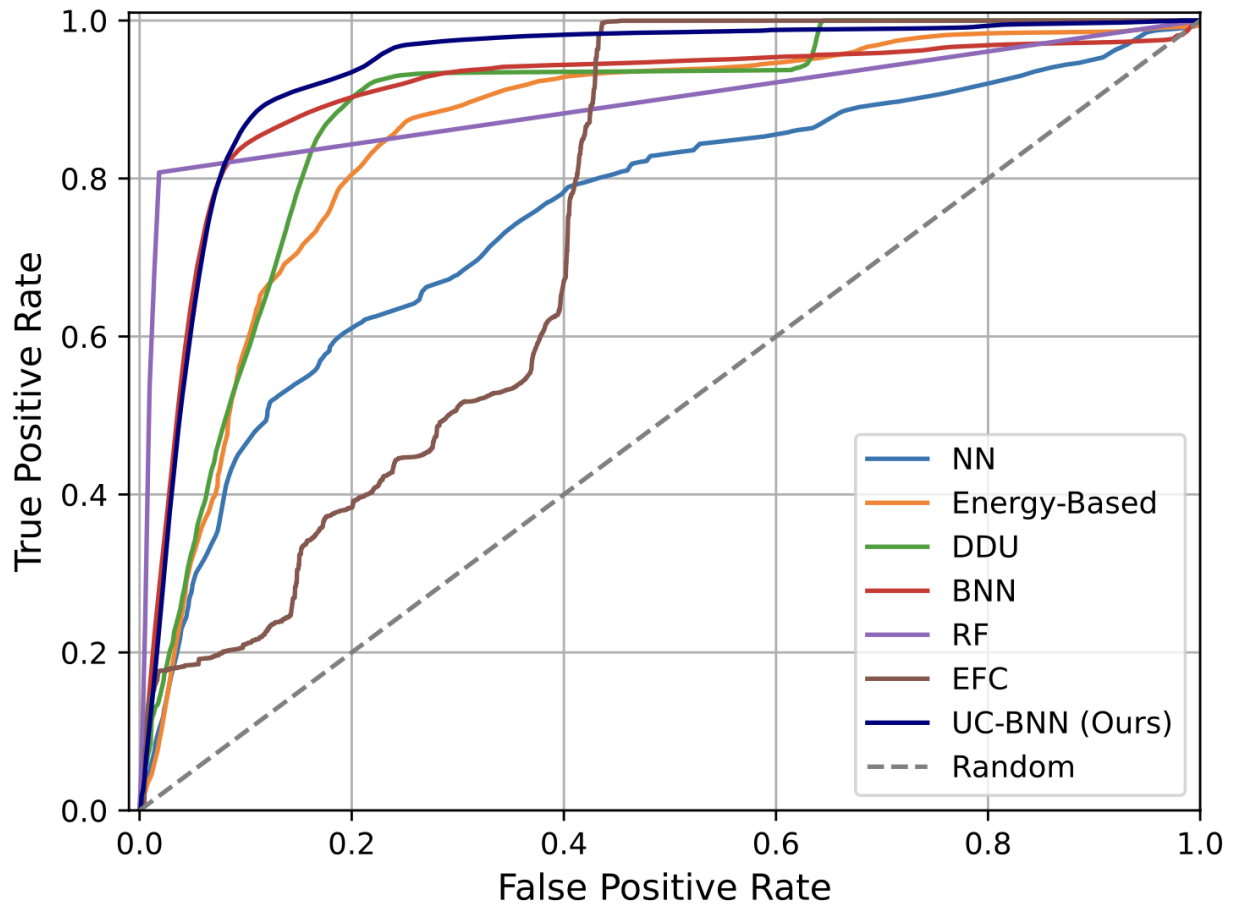

FIG. 8. Mean ROCs for the 8U scenario.

Our research reveals that conventional Neural Network-based approaches are inadequate for trustworthy network intrusion detection, due to their limited capability of quantifying predictions' uncertainty and detecting Out-of-Distribution samples. In contrast, *uncertainty-aware* methods such as BNN-based models and Deep Deterministic Uncertainty demonstrate promising potential to solve all the three problems.

Specifically, we propose a custom BNN-based model, called UC-BNN, which improves OoD detection capabilities with respect to standard BNN and stands out for its robustness, providing the most consistent results in the different OoD detection experiments, without significantly reducing the closed-set classification performance. In addition and not surprisingly, our analysis brings out uncertainty-aware Random Forests as another strong baseline for building a trustworthy ML-based IDS.

As a future work we would like to leverage *continual learning* and *class-incremental learning* to update the uncertainty-aware ML model at run time with novelties, without forgetting the previously learned classes. In addition, we would like to explore distributed learning approaches (e.g. *Federated Learning*) to build uncertainty-aware models in a collaborative way and make them fully suitable for Edge Computing operations.

## ACKNOWLEDGMENT

The research leading to these results has been partially funded by the Italian Ministry of University and Research (MUR) under the PRIN 2022 PNRR framework (EU Contribution – NextGenerationEU – M. 4,C. 2, I. 1.1), SHIELDED project, ID P2022ZWS82.

[1] Abadi, M., Barham, P., Chen, J., Chen, Z., Davis, A., Dean, J., Devin, M., Ghemawat, S., Irving, G., Isard, M., et al. (2016). Tensorflow: A system for large-scale machine learning. In *12th {USENIX} Symposium on Operating Systems Design and Implementation ({OSDI} 16)*, pages 265–283.

[2] Akiba, T., Sano, S., Yanase, T., Ohta, T., and Koyama, M. (2019). Optuna: A next-generation hyperparameter optimization framework. In *Proceedings of the 25th ACM SIGKDD International Conference on Knowledge Discovery and Data Mining*, pages 2623–2631.

[3] Ali, S., Rehman, S. U., Imran, A., Adeem, G., Iqbal, Z., and Kim, K.-I. (2022). Comparative evaluation of ai-based techniques for zero-day attacks detection. *Electronics*, 11(23):3934.

[4] Alsaedi, A., Moustafa, N., Tari, Z., Mahmood, A., and Anwar, A. (2020). Ton-iot telemetry dataset: A new generation dataset of iot and iiot for data-driven intrusion detection systems. *IEEE Access*.

[5] Apruzzese, G., Pajola, L., and Conti, M. (2022). The cross-evaluation of machine learning-based network intrusion detection systems. *IEEE Transactions on Network and Service Management*, 19(4):5152–5169.

[6] Bendale, A. and Boult, T. E. (2016). Towards open set deep networks. In *Proceedings of the IEEE conference on computer vision and pattern recognition*, pages 1563–1572.

[7] Betsy, S. W., Murugesan, A., Ganapathy, N. B. S., and Pughazendi, N. (2023). A novel framework for network intrusion detection in healthcare domain. In *2023 4th International Conference on Signal Processing and Communication (ICSPC)*, pages 43–46.

[Bishop] Bishop, C. M. *Pattern recognition and machine learning*, volume 4. Springer.

[9] Blundell, C., Cornebise, J., Kavukcuoglu, K., and Wierstra, D. (2015). Weight uncertainty in neural network. In *International conference on machine learning*, pages 1613–1622. PMLR.

[10] Bradley, A. P. (1997). The use of the area under the roc curve in the evaluation of machine learning algorithms. *Pattern recognition*, 30(7):1145–1159.

[11] Campos, E. M., Saura, P. F., González-Vidal, A., Hernández-Ramos, J. L., Bernabé, J. B., Baldini, G., and Skarmeta, A. (2022). Evaluating federated learning for intrusion detection in internet of things: Review and challenges. *Computer Networks*, 203:108661.

[12] Chen, C.-W., Su, C.-H., Lee, K.-W., and Bair, P.-H. (2020). Malware family classification using active learning by learning. In *2020 22nd International Conference on Advanced Communication Technology (ICACT)*, pages 590–595.

[13] Claise, B. (2004). Cisco systems NetFlow services export version 9. Technical report, IETF RFC.

[14] Clevert, D.-A., Unterthiner, T., and Hochreiter, S. (2015). Fast and accurate deep network learning by exponential linear units (elus). *arXiv preprint arXiv:1511.07289*.

[15] Cohn, D. A., Ghahramani, Z., and Jordan, M. I. (1996). Active learning with statistical models. *Journal of artificial intelligence research*, 4:129–145.

[16] Depeweg, S., Hernandez-Lobato, J.-M., Doshi-Velez, F., and Udluft, S. (2018). Decomposition of uncertainty in bayesian deep learning for efficient and risk-sensitive learning. In *International Conference on Machine Learning*, pages 1184–1193. PMLR.

[17] Dillon, J. V., Langmore, I., Tran, D., Brevdo, E., Vasudevan, S., Moore, D., Patton, B., Alemi, A., Hoffman, M., and Saurous, R. A. (2017). Tensorflow distributions. *arXiv preprint arXiv:1711.10604*.

[18] Doriguzzi-Corin, R., Knob, L. A. D., Mendozzi, L., Siracusa, D., and Savi, M. (2023). Introducing packet-level analysis in programmable data planes to advance network intrusion detection.

[19] European Union Agency for Cybersecurity (ENISA) (2022). ENISA Threat Landscape 2022. https://www.enisa.europa.eu/publications/enisa-threat-landscape-2022. [Accessed: 03-August-2023].

[20] Gal, Y. and Ghahramani, Z. (2016). Dropout as a bayesian approximation: Representing model uncertainty in deep learning. In *international conference on machine learning*, pages 1050–1059. PMLR.

[21] Gal, Y., Islam, R., and Ghahramani, Z. (2017). Deep bayesian active learning with image data. In *International conference on machine learning*, pages 1183–1192. PMLR.

[22] Gretton, A., Borgwardt, K. M., Rasch, M. J., Schölkopf, B., and Smola, A. (2012). A kernel two-sample test. *The Journal of Machine Learning Research*, 13(1):723–773.

[23] Guo, C., Pleiss, G., Sun, Y., and Weinberger, K. Q. (2017). On calibration of modern neural networks. In *International conference on machine learning*, pages 1321–1330. PMLR.

[24] Guo, Y. (2022). A review of machine learning-based zero-day attack detection: Challenges and future directions. *Computer Communications*.

[25] Hajizadeh, M., Barua, S., and Golchin, P. (2023). Fsa-ids: A flow-based self-active intrusion detection system. In *NOMS 2023-2023 IEEE/IFIP Network Operations and Management Symposium*, pages 1–9.

[26] Hassija, V., Chamola, V., Saxena, V., Jain, D., and et al. (2019). A survey on iot security: Application areas, security threats, and solution architectures. *IEEE Access*, 7:82721–82743.

[27] He, K., Zhang, X., Ren, S., and Sun, J. (2015). Deep residual learning for image recognition.

[28] Hendrycks, D. and Gimpel, K. (2018). A baseline for detecting misclassified and out-of-distribution examples in neural networks.

[29] Hindy, H., Atkinson, R., Tachtatzis, C., Colin, J.-N., Bayne, E., and Bellekens, X. (2020). Utilising deep learning techniques for effective zero-day attack detection. *Electronics*, 9(10):1684.

[30] Houlsby, N., Huszár, F., Ghahramani, Z., and Lengyel, M. (2011). Bayesian active learning for classification and preference learning. *arXiv preprint arXiv:1112.5745*.

[31] Ioffe, S. and Szegedy, C. (2015). Batch normalization: Accelerating deep network training by reducing internal covariate shift.

[32] Jordaney, R., Sharad, K., Dash, S. K., Wang, Z., Papini, D., Nouretdinov, I., and Cavallaro, L. (2017). Transcend: Detecting concept drift in malware classification models. In *26th USENIX Security Symposium (USENIX Security 17)*, pages 625–642.

[33] Khan, M. A., Karim, M. R., and Kim, Y. (2019). A scalable and hybrid intrusion detection system based on the convolutional-lstm network. *Symmetry*, 11(4):583.

[34] Kingma, D. P. and Ba, J. (2017). Adam: A method for stochastic optimization.

[35] Liao, H.-J., Lin, C.-H. R., Lin, Y.-C., and Tung, K.-Y. (2013). Intrusion detection system: A comprehensive review. *Journal of Network and Computer Applications*, 36(1):16–24.

[36] Liu, J., Lin, Z., Padhy, S., Tran, D., Bedrax Weiss, T., and Lakshminarayanan, B. (2020a). Simple and principled uncertainty estimation with deterministic deep learning via distance awareness. *Advances in Neural Information Processing Systems*, 33:7498–7512.

[37] Liu, J., Lin, Z., Padhy, S., Tran, D., Bedrax Weiss, T., and Lakshminarayanan, B. (2020b). Simple and principled uncertainty estimation with deterministic deep learning via distance awareness. *Advances in Neural Information Processing Systems*, 33:7498–7512.

[38] Liu, W., Wang, X., Owens, J., and Li, Y. (2020c). Energy-based out-of-distribution detection. *Advances in neural information processing systems*, 33:21464–21475.

[39] MacKay, D. J. (1995). Probable networks and plausible predictions-a review of practical bayesian methods for supervised neural networks. *Network: computation in neural systems*, 6(3):469.

[40] Min, E., Long, J., Liu, Q., Cui, J., Cai, Z., and Ma, J. (2018). Su-ids: A semi-supervised and unsupervised framework for network intrusion detection. In *Cloud Computing and Security: 4th International Conference, ICCCS 2018, Haikou, China, June 8–10, 2018, Revised Selected Papers, Part III 4*, pages 322–334. Springer.

[41] Miyato, T., Kataoka, T., Koyama, M., and Yoshida, Y. (2018). Spectral normalization for generative adversarial networks.

[42] Moustafa, N. (2021). A new distributed architecture for evaluating ai-based security systems at the edge: Network ton_iot datasets. *Sustainable Cities and Society*, 72:102994.

[43] Mukhoti, J., Kirsch, A., van Amersfoort, J., Torr, P. H., and Gal, Y. (2021). Deterministic neural networks with inductive biases capture epistemic and aleatoric uncertainty. *arXiv preprint arXiv:2102.11582*, page 13.

[44] Murphy, K. P. (2023). *Probabilistic Machine Learning: Advanced Topics*. MIT Press.

[45] Nguyen, A., Yosinski, J., and Clune, J. (2015). Deep neural networks are easily fooled: High confidence predictions for unrecognizable images. In *Proceedings of the IEEE conference on computer vision and pattern recognition*, pages 427–436.

[46] Nguyen, H. T., Yadegar, J., Kong, B., and Wei, H. (2012). Efficient batch-mode active learning of random forest. In *2012 IEEE Statistical Signal Processing Workshop (SSP)*, pages 596–599. IEEE.

[47] Nixon, J., Dusenberry, M. W., Zhang, L., Jerfel, G., and Tran, D. (2019). Measuring calibration in deep learning. In *CVPR workshops*, pages 38–41.

[48] Pedregosa, F., Varoquaux, G., Gramfort, A., Michel, V., Thirion, B., Grisel, O., Blondel, M., Prettenhofer, P., Weiss, R., Dubourg, V., Vanderplas, J., Passos, A., Cournapeau, D., Brucher, M., Perrot, M., and Duchesnay, E. (2011). Scikit-learn: Machine learning in Python. *Journal of Machine Learning Research*, 12:2825–2830.

[49] Sarhan, M., Layeghy, S., and Portmann, M. (2021). Towards a standard feature set for network intrusion detection system datasets. *Mobile Networks and Applications*, 27(1):357–370.

[50] Shaker, M. H. and Hüllermeier, E. (2020). Aleatoric and epistemic uncertainty with random forests.

[51] Shen, Y., Yun, H., Lipton, Z. C., Kronrod, Y., and Anandkumar, A. (2017). Deep active learning for named entity recognition. *arXiv preprint arXiv:1707.05928*.

[52] Shone, N., Ngoc, T. N., Phai, V. D., and Shi, Q. (2018). A deep learning approach to network intrusion detection. *IEEE Transactions on Emerging Topics in Computational Intelligence*, 2(1):41–50.

[53] Shwartz-Ziv, R., Goldblum, M., Li, Y. L., Bruss, C. B., and Wilson, A. G. (2023). On representation learning under class imbalance.

[54] Smith, F. B., Kirsch, A., Farquhar, S., Gal, Y., Foster, A., and Rainforth, T. (2023). Prediction-oriented bayesian active learning. In *International Conference on Artificial Intelligence and Statistics*, pages 7331–7348. PMLR.

[55] Smith, L. and Gal, Y. (2018). Understanding measures of uncertainty for adversarial example detection. *arXiv preprint arXiv:1803.08533*.

[56] Sommer, R. and Paxson, V. (2010). Outside the closed world: On using machine learning for network intrusion detection. In *2010 IEEE Symposium on Security and Privacy*, pages 305–316.

[57] Souza, M. M. C., Pontes, C., Gondim, J., Garcia, L. P. F., DaSilva, L., and Marotta, M. A. (2022). A novel open set energy-based flow classifier for network intrusion detection.

[58] Tauscher, Z., Jiang, Y., Zhang, K., Wang, J., and Song, H. (2021). Learning to detect: A data-driven approach for network intrusion detection. In *2021 IEEE International Performance, Computing, and Communications Conference (IPCCC)*, pages 1–6. IEEE.

[59] Tsimenidids, S., Lagkas, T., and Rantos, K. (2022). Deep learning in iot intrusion detection. *Journal of Network and Systems Management*, 30.

[60] Van Amersfoort, J., Smith, L., Teh, Y. W., and Gal, Y. (2020a). Uncertainty estimation using a single deep deterministic neural network. In *International conference on machine learning*, pages 9690–9700. PMLR.

[61] Van Amersfoort, J., Smith, L., Teh, Y. W., and Gal, Y. (2020b). Uncertainty estimation using a single deep deterministic neural network. In *International conference on machine learning*, pages 9690–9700. PMLR.

[62] Varghese, B., Wang, N., Barbhuiya, S., Kilpatrick, P., and Nikolopoulos, D. S. (2016). Challenges and opportunities in edge computing. In *2016 IEEE International Conference on Smart Cloud (SmartCloud)*, pages 20–26.

[63] Verma, A. and Ranga, V. (2020). Machine learning based intrusion detection systems for iot applications. *Wireless Personal Communications*, 111:2287–2310.

[64] Vinayakumar, R., Alazab, M., Soman, K., Poornachandran, P., Al-Nemrat, A., and Venkatraman, S. (2019). Deep learning approach for intelligent intrusion detection system. *Ieee Access*, 7:41525–41550.

[65] Wen, Y., Vicol, P., Ba, J., Tran, D., and Grosse, R. (2018). Flipout: Efficient pseudo-independent weight perturbations on mini-batches.

[66] Wilson, A. G. and Izmailov, P. (2020). Bayesian deep learning and a probabilistic perspective of generalization. *Advances in neural information processing systems*,

33:4697–4708.

[67] Yang, J., Zhou, K., Li, Y., and Liu, Z. (2021). Generalized out-of-distribution detection: A survey. *arXiv preprint arXiv:2110.11334*.

[68] Ye, Y., Zhang, T., and Yang, C. (2019). Fisher loss: A more discriminative feature learning method in classification. In *2019 IEEE/ASME International Conference on Advanced Intelligent Mechatronics (AIM)*, pages 746–751.

[69] Zhang, Z., Zhang, Y., Guo, D., and Song, M. (2021). A scalable network intrusion detection system towards detecting, discovering, and learning unknown attacks. *International Journal of Machine Learning and Cybernetics*, 12:1649–1665.

[70] Zoppi, T., Ceccarelli, A., and Bondavalli, A. (2021). Unsupervised algorithms to detect zero-day attacks: Strategy and application. *IEEE Access*, 9:90603–90615.